\pdfoutput=1
\documentclass{vldb}
\usepackage{booktabs} 

\usepackage{subfig}
\usepackage{wrapfig}
\usepackage{lipsum}
\usepackage{graphicx}
\usepackage{multirow}
\usepackage{url}

\usepackage[ruled, vlined, linesnumbered]{algorithm2e}

\usepackage[table,usenames, dvipsnames]{xcolor}

\newtheorem{property}{Property}

\usepackage{soul,xspace}

\newcommand{\moursyst}{\text{Nuri}}
\newcommand{\oursyst}{$\moursyst$\xspace}

\usepackage{listings}
\usepackage{textcomp}
\newcommand{\agg}{\mathcal{G}}

\usepackage{graphicx}
\usepackage{balance}  

\begin{document}

\title{An Efficient System for Subgraph Discovery}

\numberofauthors{1} 

\author{
%
%
\alignauthor
Aparna Joshi, Yu Zhang, Petko Bogdanov, and Jeong-Hyon Hwang\\
       \affaddr{Department of Computer Science}\\
       \affaddr{University at Albany -- State University of New York}\\
       \affaddr{1400 Washington Avenue, Albany, NY 12222, USA}\\
       \affaddr{\{ajoshi, yzhang20, pbogdanov, jhwang\}@albany.edu}
}

\maketitle


\begin{abstract}

Subgraph discovery in a single data graph---finding subsets of vertices and edges satisfying a user-specified criteria---is an essential and general graph analytics operation with a wide spectrum of applications. Depending on the criteria, subgraphs of interest may correspond to cliques of friends in social networks, interconnected entities in RDF data, or frequent patterns in protein interaction networks to name a few. Existing systems usually examine a large number of subgraphs while employing many computers and often produce an enormous result set of subgraphs. How can we enable fast discovery of only the most relevant subgraphs while minimizing the computational requirements?   

We present Nuri, a general subgraph discovery system that allows users to succinctly specify subgraphs of interest and criteria for ranking them.
Given such specifications, Nuri efficiently finds the $k$ most relevant subgraphs using only a single computer.
It prioritizes (i.e., expands earlier than others) subgraphs that are more likely to expand into the desired subgraphs (\emph{prioritized subgraph expansion}) and proactively discards irrelevant subgraphs from which the desired subgraphs cannot be constructed (\emph{pruning}). 
Nuri can also efficiently store and retrieve a large number of subgraphs on disk without being limited by the size of main memory.
We demonstrate using both real and synthetic datasets that Nuri on a single core outperforms the closest alternative distributed system consuming 40 times more computational resources by more than 2 orders of magnitude for clique discovery and 1 order of magnitude for subgraph isomorphism and pattern mining.

\end{abstract}



\section{Introduction}
\label{sec:introduction}
Social networks, the World Wide Web, transportation networks and protein-protein interaction (PPI) networks are commonly modeled as {\em graphs} in which {\em vertices} represent entities, {\em edges} represent relationships between entities, and vertex/edge labels represent certain properties of the entities/relationships.
Given such a graph, the problem of finding {\em subgraphs} (subsets of vertices and edges) that meet specific user-defined criteria arises in a variety of applications.
For example, prominent star and clique structures of high homophily (i.e., sharing attributes) in social networks have been shown instrumental to understanding the nature of society~\cite{holder2005graph,chau2007mining,coffman2004graph,jiang2013survey}.
In the biological domain, subgraphs in a PPI network of highest gene expression disagreement across phenotypes (e.g., healthy/sick) are essential for identifying target pathways and complexes to manipulate a condition~\cite{ranu2013mining,hu2005mining,li2010computational,Dang2014,DangAndYou2015}.
Other applications include computer network security~\cite{zhao2009botgraph,nagaraja2010botgrep,coskun2010friends}, financial fraud detection~\cite{eberle2010insider,phua2010comprehensive}, and community discovery in social and collaboration networks~\cite{papadopoulos2012community,plantie2013survey}.

In response to the aforementioned demand, researchers have developed custom solutions for specific types of subgraph discovery problems.
Examples include subgraph isomorphism 
search algorithms~\cite{han.sigmod13.turboiso,lee.pvldb12.comparison,gupta.icde14.topksubgraph,ullmann.j1976.iso} and techniques for discovering frequent subgraphs~\cite{elseidy.pvldb14.grami, zhu.pvldb11.topkfsm, anchuri.sigkdd13.graphming,garey.wh02.book}, cliques~\cite{carraghan.orl1990.clique, wu.ejor15.cliquereview,bron.commun73.clique,uno,eppstein.sea11.clique,cheng.sigkdd12.cliqueMPI}, quasi-cliques and dense subgraphs ~\cite{Andersen2008}, as well as communities~\cite{raghavan2007near,yang2015defining}. 
The above techniques, however, are one-off solutions for specific problems, do not provide the necessary system support for large-scale computations, and are usually difficult to use/extend for different subgraph discovery computations.

Various types of graph computations including PageRank~\cite{brin.computerNetworks98.google.pagerank} have been supported by TLV (``think like a vertex'') graph systems including Pregel~\cite{malewicz.sigmod10.pregel,ching.hadoopSummit11.giraph}, GraphLab~\cite{low.uai10.graphlab},
Graph- -Chi~\cite{kyrola.osdi12.graphchi}, and TurboGraph~\cite{han.sigkdd13.turbograph}.
These systems, which store the state of computation in the form of vertex attributes, however, are not suitable for subgraph discovery computations.
The reason is that subgraphs cannot be adequately expressed as vertex attributes since the number of subgraphs grows exponentially with the size of the data graph and there is typically a many-to-many relationship between vertices and subgraphs of interest.

Different from the above, systems specifically targeted to subgraph discovery have recently been proposed~\cite{teixeira.sosp15.arabesque, quamar2014nscale, chen2018g}. 
These systems adopt {\em subgraph exploration} as the building block for subgraph discovery computations.
In other words, they initially construct one-vertex (or one-edge) subgraphs and then repeatedly expand subgraphs into larger ones by adding a vertex or an edge at a time.
As discussed later in this paper, however, these systems often exhibit limited performance (even when they employ a large number of servers) mainly due to the sheer number of subgraphs that they have to examine.
They may also produce an overwhelmingly large result set, for example, millions of subgraphs, which cannot be easily dealt with by a human analyst.

We propose a new subgraph discovery system, called \oursyst, that overcomes the above limitations.
\oursyst supports a variety of subgraph discovery computations (i) {\em conveniently} (as opposed to the complexity of developing custom solutions~\cite{zou2007top,wu2013ontology,yang2016fast,hong2015subgraph,gupta.icde14.topksubgraph,vasilyeva2014top,macropol2010scalable}, for example, managing a large number of subgraphs that cannot fit into the main memory) via an API that enables succinct implementation of these computations and (ii) more {\em efficiently} compared to the closest existing alternative systems~\cite{teixeira.sosp15.arabesque, quamar2014nscale, chen2018g} by quickly finding the $k$ most relevant subgraphs according to user-provided specifications. 
The key advantageous features of Nuri are as follows:

\noindent
{\bf  Targeted Expansion.}
Our API allows users to specify whether or not it is adequate to expand a subgraph by adding a vertex or an edge.
This feature enables \oursyst to create and examine only the necessary subgraphs in contrast to Arabesque~\cite{teixeira.sosp15.arabesque} which exhaustively creates subgraphs and then filters out irrelevant ones.

\noindent
{\bf  Prioritization.}
Our API allows users to assign a higher priority to subgraphs that are more likely to expand into subgraphs of interest than others (e.g., cliques that have the potential to expand into larger cliques).
Given this specification, \oursyst expands subgraphs with a higher priority before other subgraphs, thereby speeding up the discovery of the desired subgraphs.
All of the previous subgraph discovery systems ~\cite{teixeira.sosp15.arabesque, quamar2014nscale, chen2018g} lack this feature and thus have inherent performance limitation.

\noindent
{\bf Pruning.}
As soon as the result set contains $k$ entries, it becomes unnecessary to expand subgraphs whose expansions cannot lead to subgraphs that are more relevant than the $k$ entries (e.g., in the case of finding the largest cliques, cliques that can only expand into cliques of up to size 3 when the result set already contains a clique of size 4).
\oursyst can detect and safely discard such subgraphs according to a user specification.

\noindent
{\bf Efficient Top-$\boldsymbol{k}$ Aggregate Subgraph Discovery.}
While some key subgraph discovery computations require grouping of subgraphs (e.g., by their patterns) and aggregation of certain properties (e.g., calculation of pattern frequency)~\cite{elseidy.pvldb14.grami, zhu.pvldb11.topkfsm, anchuri.sigkdd13.graphming,garey.wh02.book}, previous subgraph discovery systems such as NScale~\cite{quamar2014nscale} and GMiner~\cite{chen2018g} cannot support such {\em aggregate} computations.
In contrast to Arabesque~\cite{teixeira.sosp15.arabesque} that must expand all smaller subgraphs before any larger subgraph, \oursyst can expand a group of subgraphs before other smaller subgraphs ({\em prioritized expansion}) thereby facilitating {\em early} and {\em effective pruning}.

\noindent
{\bf On-Disk Subgraph Management.}
The number of subgraphs that \oursyst manages usually grows exponentially with the size and density of the data graph and may even surpass the capacity of the main memory.
\oursyst has the ability to manage high-priority subgraphs in memory and low-priory subgraphs on disk in a highly efficient manner where the use of disk causes only a slight degradation in performance.

In summary, this paper presents the following contributions by us:

\begin{itemize}

\item two new computational models that efficiently support various top-$k$ subgraph discovery computations through prioritized subgraph expansion and pruning

\item an API that allows users to easily implement diverse subgraph discovery computations (and examples demonstrating succinct implementation of representative subgraph discovery algorithms)

\item design and implementation of a system, \oursyst, that enables fast top-$k$ subgraph discovery just on a single computer

\item in-depth analysis of experimental results which demonstrate between $1$ to $2$ orders of magnitude reduction of subgraph discovery time for our system running on a single computer, compared to the closest alternative distributed system utilizing $40$ times more computational resources.

\end{itemize}

The rest of this paper is organized as follows: 
Section~\ref{sec:background} introduces example subgraph discovery computations and discusses previous systems that are most closely related to our work.
Section~\ref{sec:model} presents our computational models which enable fast subgraph discovery through prioritized subgraph expansion and pruning.
Sections~\ref{sec:api} and \ref{sec:architecture} 
 explain our API and system architecture, respectively. 
Section~\ref{sec:evaluation} explains our evaluation results, Section~\ref{sec:related_work} summarizes related work, and Section~\ref{sec:conclusion} concludes this paper.


\section{Background}
\label{sec:background}

In this section, we discuss several popular subgraph discovery computations that we adopt to describe and evaluate our general system (Section~\ref{subsec:queries}).
We also explain previous systems/solutions that are closely related to our work and their limitations (Section~\ref{subsec:limitations}). 

\subsection{Subgraph Discovery Computations}
\label{subsec:queries}

Our goal is to enable efficient discovery of the most relevant subgraphs that meet user-specified criteria within a large data graph. For demonstrative purposes, we adopt the following example discovery computations:

\begin{figure}[t]
    \begin{center}
     \subfloat[Graph with 4 vertices\label{fig:cliques_graph}]{%
       \includegraphics[width=0.11 \textwidth, trim={-.1cm .01cm -.1cm 0}]{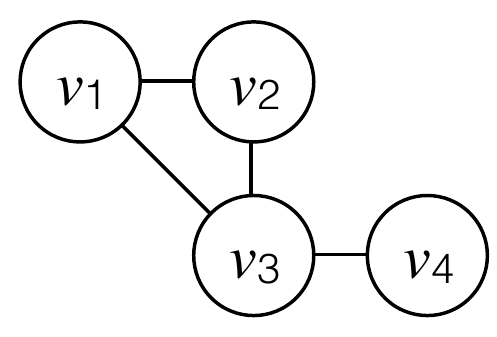}
     }
     \hfill
     \subfloat[Graph with 5 vertices ($a$ and $b$ are labels)\label{fig:patterns_graph}]{%
       \includegraphics[width=0.19\textwidth, trim={-0.5cm .01cm -0.5cm 0}]{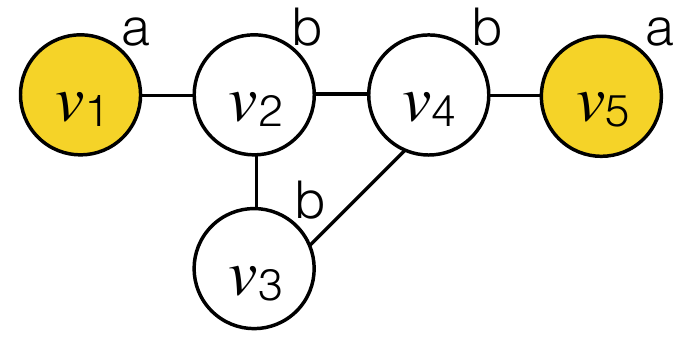}
     }
     \hfill
     \subfloat[Query/ pattern graph\label{fig:iso_query}]{%
       \includegraphics[width=0.13\textwidth, trim={-.4cm -.9cm -.4cm 0}]{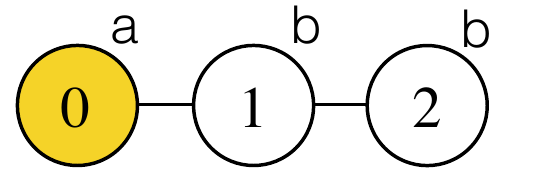}
     }
    \end{center}
    \caption{
    Sample Graphs}\label{fig:sample_graphs}
\end{figure}

\begin{figure}[t]
    \includegraphics[width=.46\textwidth, trim={0 .2cm 0 0}]{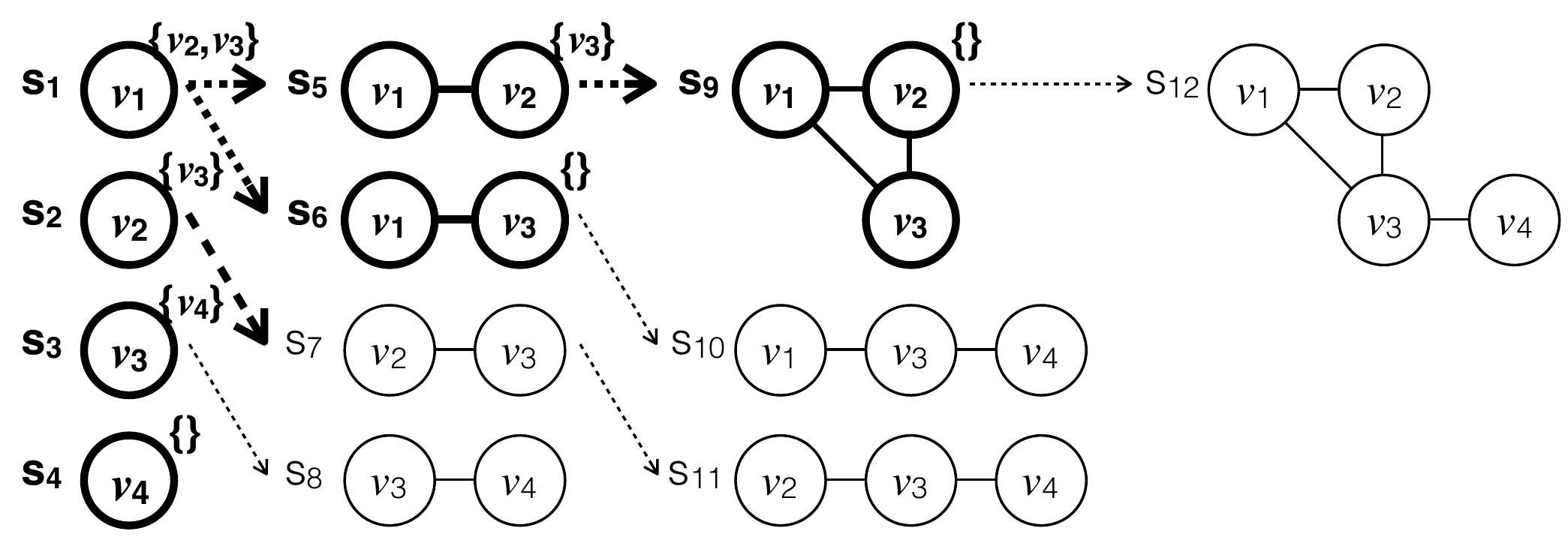}
    \caption{
    Clique Discovery (on the graph in Figure~\ref{fig:cliques_graph})
    }
    \label{fig:cliques_process}
\end{figure}

\noindent
{\bf Clique Discovery.} 
A clique is a subgraph in which every pair of vertices are adjacent~\cite{carraghan.orl1990.clique, wu.ejor15.cliquereview,bron.commun73.clique,uno,eppstein.sea11.clique,cheng.sigkdd12.cliqueMPI}.
The data graph in Figure~\ref{fig:cliques_graph} contains $9$ cliques of sizes $1$ to $3$, depicted as $s_1$, $s_2$, $s_3$, $s_4$, $s_5$, $s_6$, $s_7$, $s_8$, and $s_9$ in Figure~\ref{fig:cliques_process} (details of this figure are explained in Sections~\ref{subsec:limitations} and \ref{subsec:non-aggregate}). 

\begin{figure}[t]
    \includegraphics[width=.48\textwidth, trim={0 .2cm 0 0}]{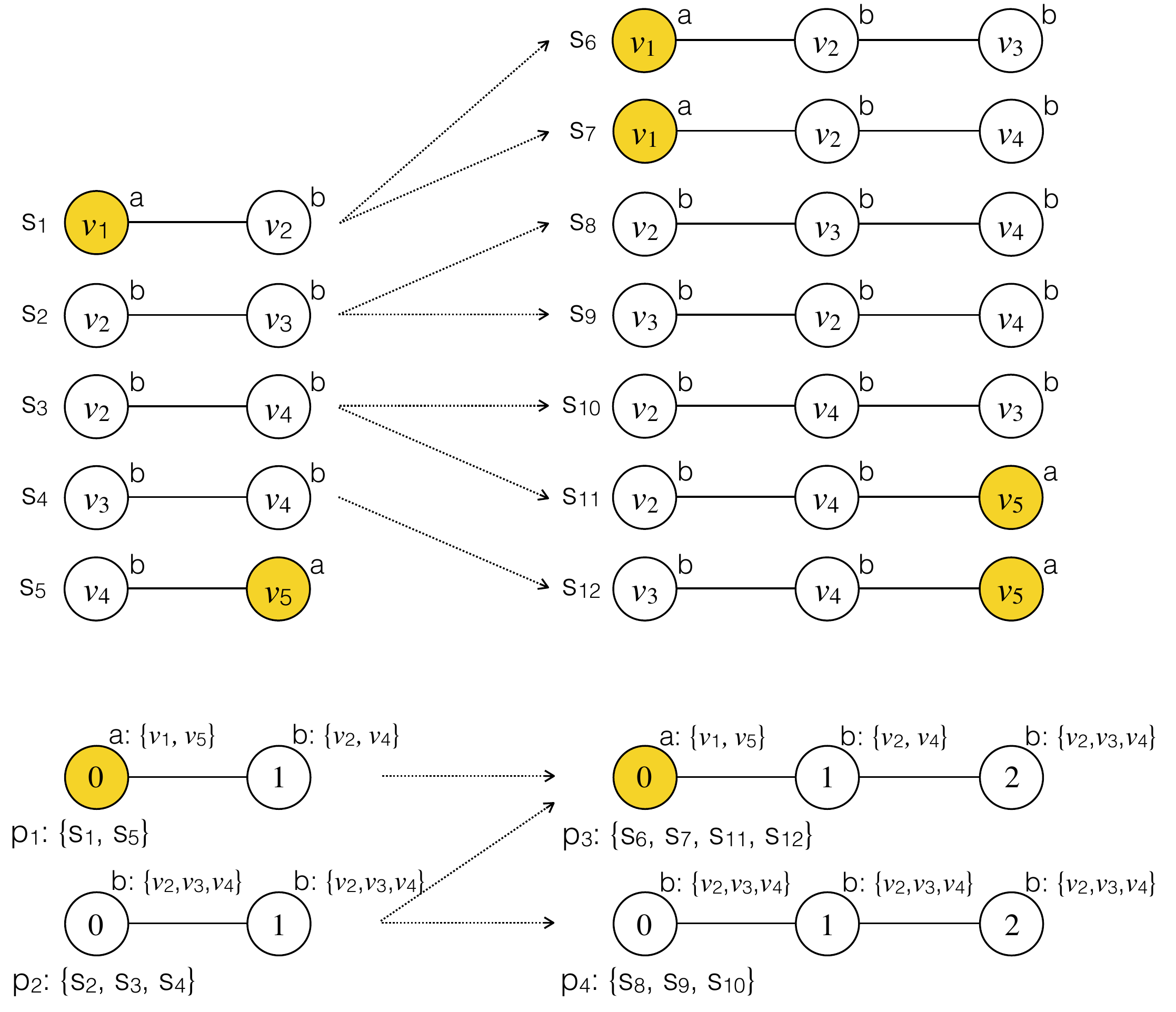}
    \caption{Pattern Mining (on the graph in Figure~\ref{fig:patterns_graph})
    }\label{fig:patterns_process}
\end{figure}

\noindent{\bf Subgraph Isomorphism.}
The goal of subgraph isomorphism search is to find all subgraphs isomorphic to a query graph 
in a labeled data graph~\cite{han.sigmod13.turboiso,lee.pvldb12.comparison,gupta.icde14.topksubgraph,ullmann.j1976.iso}.
For a query graph $G_q(V_q, E_q, L_q)$, a subgraph $G_s(V_s, E_s, L_s)$ in the data graph is isomorphic to the query graph if there exists a bijection $\mu:V_q \rightarrow V_s$ such that (1) $\forall{v \in V_q}, L_q(v) = L_s(\mu(v))$ (i.e., the label of vertex $v$ in $G_q$ is the same as the label of the corresponding vertex $\mu(v)$ in $G_s$) and (2) $\forall{(v, v') \in E_q} \Leftrightarrow (\mu(v), \mu(v')) \in E_s$ (i.e., vertices $v$ and $v'$ are adjacent in $G_q$ if and only if the corresponding vertices $\mu(v)$ and $\mu(v')$ are adjacent in $G_s$).
For example, the data graph in Figure~\ref{fig:patterns_graph} contains four subgraphs (depicted as $s_6$, $s_7$, $s_{11}$, and $s_{12}$ in Figure~\ref{fig:patterns_process}) that are isomorphic to the query graph in Figure~\ref{fig:iso_query}.
Subgraph $s_6$ is isomorphic to the query graph since a bijection such that $\mu(0)=v_1$,  $\mu(1)=v_2$,  and $\mu(2)=v_3$ satisfies the conditions mentioned above.

\noindent{\bf Pattern Mining.} 
The goal of pattern mining is to find subgraph patterns that appear at least as frequently as a user-specified threshold in the data graph. 
For example, the subgraph pattern in Figure~\ref{fig:iso_query} appears in the data graph from Figure~\ref{fig:patterns_graph} (the subgraphs depicted as $s_6$, $s_7$,  $s_{11}$, and $s_{12}$ in Figure~\ref{fig:patterns_process} are isomorphic to the pattern in Figure~\ref{fig:iso_query}).

Among several pattern frequency definitions~\cite{fiedler.icdm07.ho,kuramochi.dmkd05.mis,bringmann.pakdd08.minimImageBased}, for the purposes of this example, we consider the {\em minimum image-based support}, which is defined as the minimum number of mappings for any vertex in the pattern to the corresponding vertices in the data graph~\cite{bringmann.pakdd08.minimImageBased}.
According to this definition, the frequency of pattern $p_1$ in Figure~\ref{fig:patterns_process} is 2 since (i) subgraphs $s_1$ and $s_5$ match (i.e., are isomorphic to) $p_1$ and (ii) the first vertex of $p_1$ (vertex $0$) maps to vertices $v_1$ and $v_5$ in Figure~\ref{fig:patterns_graph}, (iii) the second vertex of $p_1$ (vertex $1$) maps to vertices $v_2$ and $v_4$, and thus (iv) the frequency of $p_1$, denoted $f(p_1)$, is $\min(|\{v_1, v_5\}|, |\{v_2, v_4\}|) = \min(2, 2) = 2$.
The other patterns in Figure~\ref{fig:patterns_process} have the following frequencies: $f(p_2) =  \min(|\{v_2, v_3, v_4\}|, |\{v_2, v_3, v_4\}|) = \min(3, 3) = 3$, $f(p_3) = \min(|\{v_1, v_5\}|, |\{v_2, v_4\}|, |\{v_2, v_3, v_4\}|) = \min(2, 2, 3) = 2$, and $f(p_4) = \min(|\{v_2, v_3, v_4\}|, |\{v_2, v_3, v_4\}|, |\{v_2, v_3, v_4\}|) = \min(3, 3, 3) = 3$.

\noindent{\bf Top-$\boldsymbol{k}$ Semantics.} 
Given a large graph, the above computations may return an enormous number of subgraphs, thus overwhelming the user. 
Our goal is to allow users to specify the desired size $k$ of the result set and a criterion for ranking subgraphs.
Examples include (i) the cliques of the largest size, (ii) the isomorphic matches of the highest cumulative edge weight, and (iii) the $k$ most frequent patterns of a certain size. 
The system then should obtain such top-$k$ results more efficiently than exhaustively acquiring all results and ranking them a posteriori.   

\subsection{Limitations of Current Graph Systems} 
\label{subsec:limitations}

Graph processing systems such as 
Pregel~\cite{malewicz.sigmod10.pregel},
Giraph~\cite{ching.hadoopSummit11.giraph}, GraphLab~\cite{low.uai10.graphlab, gonzalez.osdi12.powergraph}, GraphChi~\cite{kyrola.osdi12.graphchi}, TurboGraph~\cite{han.sigkdd13.turbograph} and 
others~\cite{xin.grades13.graphx,
salihoglu.ssdbm13.gps, shao.sigmod13.trinity, wang.atc15.graphq, wang.sigmod16.hybridgraph} adopt
a ``think like a vertex (TLV)'' computational model.
These systems iteratively update the state (i.e., variables) of each vertex in a manner that eventually computes quantities of interest, such as  PageRank~\cite{brin.computerNetworks98.google.pagerank}.
Subgraph discovery computations, however, cannot be succinctly expressed using vertex variables since the number of subgraphs grows exponentially with the size of the graph and there is typically a many-to-many relationship between vertices and subgraphs.
For this reason, TLV systems cannot adequately support subgraph discovery computations.

In contrast to the above ones, several systems specifically targeted to subgraph discovery have recently been proposed~\cite{teixeira.sosp15.arabesque, quamar2014nscale, chen2018g}. 
These systems adopt {\em subgraph exploration} as the building block for subgraph discovery computations.
In other words, they initially construct one-vertex subgraphs (e.g., $s_1$, $s_2$, $s_3$, and $s_4$ in Figure~\ref{fig:cliques_process}) or one-edge subraphs (e.g., $s_1$, $s_2$, $s_3$, $s_4$, and $s_5$ in Figure~\ref{fig:patterns_process}) and then expand subgraphs into larger ones by adding a vertex or an edge at a time (e.g., in Figure~\ref{fig:patterns_process}, $s_1$ into $s_6$ by adding edge $\{v_2, v_3\}$).

Among the above systems, only Arabesque~\cite{teixeira.sosp15.arabesque} can support  pattern mining 
by initially constructing one-edge subgraphs and then grouping these subgraphs according to their patterns (e.g., in Figure~\ref{fig:patterns_process}, a group of subgraphs $s_1$ and $s_5$ that match pattern $p_1$ and another group of subgraphs $s_2$, $s_3$, and $s_4$ that match pattern $p_2$). 
It then computes the frequency of each pattern and repeats the process of expanding, for each pattern whose frequency is no lower than a threshold, the subgraphs that match the pattern.

The limitations of the above systems are as follows: First, as further explained in Section~\ref{subsec:non-aggregate}, they cannot perform {\em prioritized expansion} of subgraphs according to user-specified criteria, inherently losing the opportunity to quickly fill the result set and start pruning out irrelevant subgraphs whose expansions cannot affect the result set.
Second, NScale~\cite{quamar2014nscale} and GMiner~\cite{chen2018g} cannot support aggregate computations such as frequent pattern mining.
Third, Arabesque~\cite{teixeira.sosp15.arabesque} adopts {\em exhaustive expansion} (i.e., constructs all subgraphs that can be obtained by adding a vertex/edge to an existing subgraph) and {\em post-expansion filtering} (i.e., discards irrelevant subgraphs), and thus may create a large number of unnecessary subgraphs (e.g., non-clique subgraphs  such as $s_{10}$, $s_{11}$, and $s_{12}$ in Figure~\ref{fig:cliques_process}).

\begin{table*}[t]
\small
\caption{User Functions}
\label{table:functions}
\begin{tabular}{*{4}{| l | c | p{11cm} | l | l}}
\hline
Function & Optional & Description & Default\\ 
\hline
$expandable(s, \delta)$ 
& yes & returns \texttt{true} if it is adequate to expand subgraph $s$ by adding a vertex or an edge $\delta$ & returns \texttt{true}\\ \hline
$relevant(s)$ & no &returns \texttt{true} if subgraph $s$ matches the user's interest & N/A \\ \hline
$priority(s)$  & yes &returns the application-specific priority of subgraph $s$
& returns \texttt{null}\\ 
\hline
$dominated(s, s')$ & yes & returns \texttt{true} if all subgraphs into which $s$ can expand are guaranteed to have a lower priority than subgraph $s'$ & returns \texttt{false}\\
\hline
\end{tabular}
\end{table*}

\section{Computational Model}
\label{sec:model}

In this section, we explain the basic principles of our computational model (Section~\ref{subsec:core_ideas}), its technical details (Section~\ref{subsec:non-aggregate}), and an extension to the model for supporting aggregate computations (Section~\ref{subsec:aggregate}). 

\subsection{Core Ideas}
\label{subsec:core_ideas}

Our computational model aims to quickly find the $k$-most relevant results according to user-specified preference functions.
Its key principles are as follows:
\begin{itemize}

\item {\em Targeted expansion}: our computational model initially creates unit subgraphs that contain a vertex (or an edge) and then obtains new subgraphs by repeatedly adding vertices and edges to existing subgraphs. It allows users to specify whether or not it is adequate to expand subgraph $s$ by adding a vertex or an edge $\delta$ (see the $expandable(s, \delta)$ function in Table~\ref{table:functions}). 
For example, users can avoid creation of non-clique subgraphs by making $expandable(s, \delta)$ return \texttt{true} only if adding $\delta$ to $s$ maintains a clique subgraph.

\item {\em Result ranking}: by implementing the $relevant(s)$ function (Table~\ref{table:functions}), users can specify whether or not subgraph $s$ matches their interest (e.g., $s$ is a clique) and thus may be added to the result set. 
Users can also incorporate their criteria for ranking results into the $priority(s)$ function (e.g., they can express preference for larger cliques by assigning a higher priority value to larger cliques).

\item {\em Pruning}: as explained in Section~\ref{subsec:non-aggregate}, it may be possible to calculate an {\em upper bound} on the priorities of all possible subgraphs into which a subgraph $s$ can expand (e.g., expansions of $s$ would only lead to cliques of size $3$ or less). 
In this case, users can instrument the $dominated(s, s')$ function to return \texttt{true} if all supergraphs that $s$ can expand into are guaranteed to have a lower priority than $s'$.
When $s'$ is the $k$-th entry in the result set and $dominated(s, s')$ returns \texttt{true}, it is safe to ignore subgraph $s$ since expansions of $s$ can never affect the result set (i.e., all of the subgraphs obtained through these expansions would have a lower priority than $s'$ and thus never be included in the result set).

\item {\em Prioritized expansion}: our model expands highest priority subgraphs first.
This feature allows users to effectively facilitate pruning by implementing the $priority(s)$ function (further details explained in Section~\ref{subsec:non-aggregate}).


\end{itemize}

The advantages of our approach compared to prior subgraph discovery systems~\cite{teixeira.sosp15.arabesque, quamar2014nscale, chen2018g} are discussed at the end of Sections~\ref{subsec:non-aggregate} and \ref{subsec:aggregate}.

\subsection{Basic Computational Model}
\label{subsec:non-aggregate}

\begin{algorithm}[h]
\small
\SetKw{And}{and}
\SetKw{Or}{or}
\For {$\textnormal{vertex}~v: V$ \textnormal{or} \textnormal{edge}~$e: E$} {
    create a subgraph $s$ containing only vertex $v$ or edge $e$;\\
    insert $s$ into $Q$;\\
}
\While {$|Q| > 0$} {
    subgraph $s\leftarrow remove\_max(Q)$;\\
    \If{$relevant(s)$~\And\\ $~~~(|R| < k~\Or~priority(s) \ge priority(k\textnormal{-}th(R)))$} {
        insert $s$ into $R$;\\
        \If{$|R| > k$}{
            remove from $R$ each entry $\epsilon$ such that $priority(\epsilon) < priority(k\textnormal{-}th(R))$;\\       
        }
    }
    \If{$|R| < k~\Or~!dominated(s,k\textnormal{-}th(R))$}{
        \For {each neighboring vertex (or edge) $\delta$ of  $s$} {
            \If{$expandable(s,\delta)$}{ 
                 create subgraph $s'$  by adding $\delta$ to $s$;\\
                \If{$|R| < k~\Or~!dominated(s',k\textnormal{-}th(R))$} {
                    insert $s'$ into $Q$;\\
                }
            }
        }
    }
}

\caption{Basic Computational Model}\label{alg:whole_process_noAgg}
\end{algorithm}

Algorithm~\ref{alg:whole_process_noAgg} illustrates how our computational framework carries out subgraph discovery computations.
It first creates, for each vertex (or edge) in the data graph, a subgraph $s$ containing that vertex (or edge) and inserts $s$ into a priority queue $Q$ (lines 1-3).
Next, as long as $Q$ contains subgraphs (line 4), it repeatedly dequeues and processes the subgraph $s$ with the highest priority (lines 5-16).
If $s$ matches the user's interest (line 6) and if the result set $R$ is not yet full or the priority of $s$ exceeds the $k$-th entry in $R$ (line 7), it adds $s$ to $R$ (line 8) and removes unnecessary entries from $R$ (lines 9 and 10).
Furthermore, it examines if subgraph $s$ can be safely ignored (i.e., unnecessary to expand $s$) (line 11).
If not, it considers each neighboring vertex (or edge) $\delta$ of $s$ (line 12).
If adding $\delta$ to $s$ is adequate (line 13; e.g., this expansion will lead to a clique), it expands $s$ into $s'$ by adding $\delta$ (line 14).
If $s'$ cannot be ignored (line 15), it inserts $s'$ into $Q$ (line 16).

\begin{figure}[t]
    \includegraphics[width=.49\textwidth, trim={0 .5cm 0 0}]{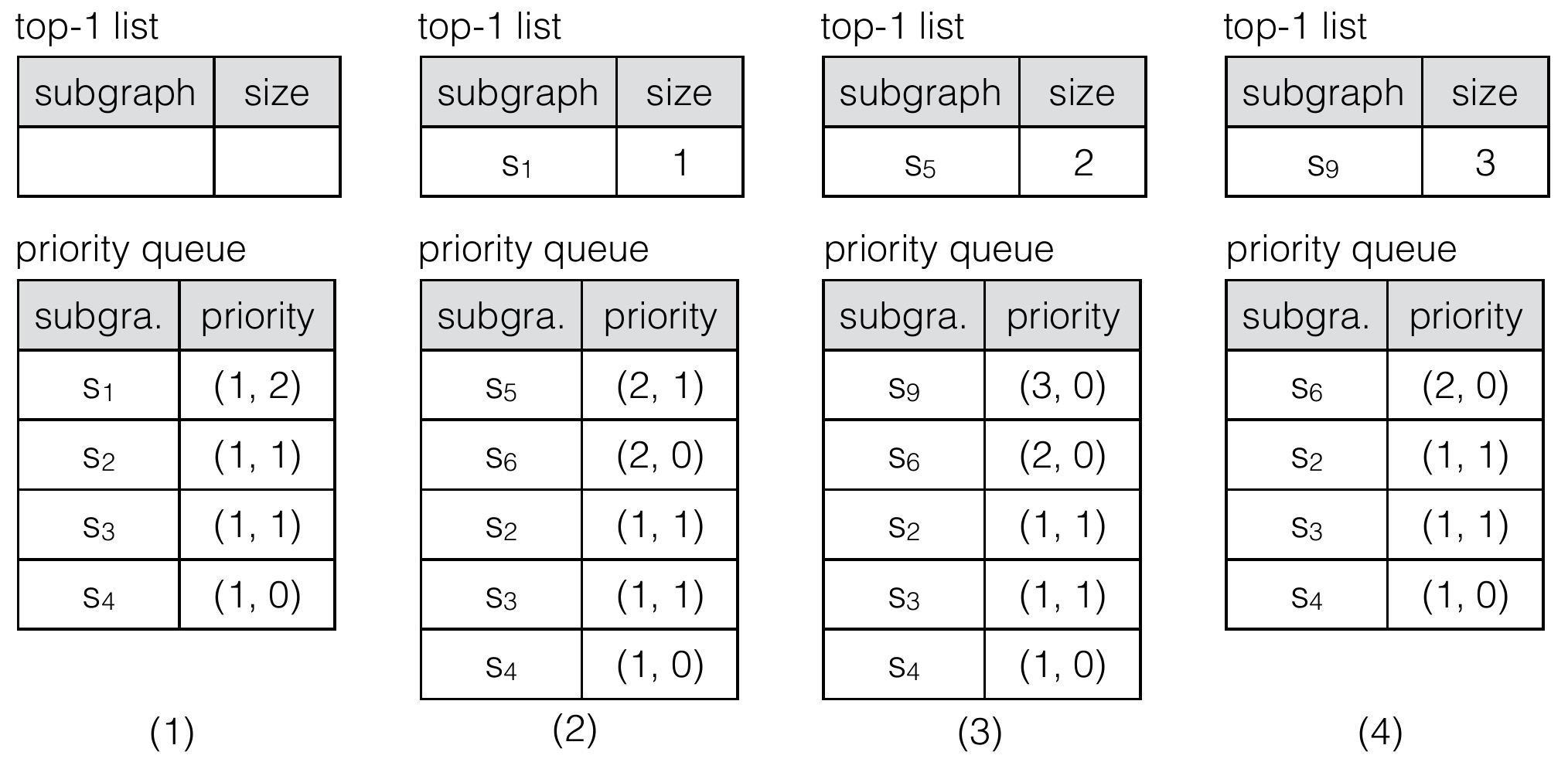}
   \caption{Maximum Clique Discovery}\label{fig:cliques_prioritized}
\end{figure}

The following example illustrates how our computational framework can efficiently support a subgraph discovery computation by applying a classical algorithm~\cite{carraghan.orl1990.clique}.


\noindent
{\bf Example: Maximum Clique Discovery.}
Assume that a user wants to quickly find the maximum clique(s) given a data graph in Figure~\ref{fig:cliques_graph}.
For each clique $s$, the user can consider the set $P_s$ of vertices that may be added to $s$ while forming a clique~\cite{carraghan.orl1990.clique} (for details of $P_s$, refer to Listing~\ref{listing:cliques}).
For example, in Figure~\ref{fig:cliques_process}, $P_{s_1} = \{ v_2, v_3\}$ since clique $s_1$ has two neighboring vertices ($v_2$ and $v_3$) whose addition (as well as their edges to a vertex in $s$) \emph{may} expand $s_1$ into a clique of size 3.
On the other hand, $P_{s_2} = \{v_3\}$ because $s_2$ has only one such neighboring vertex ($v_3$) (note that, just like Arabesque~\cite{teixeira.sosp15.arabesque}, our framework does not consider adding vertex $v_1$ to $s_2$ since this expansion would result in a duplicate generation of $s_5$ which is to be obtained by adding vertex $v_2$ to $s_1$).
The user can instrument our framework as follows (for the actual implementation of the custom functions, refer to Section~\ref{subsec:api_noAgg}):
\begin{itemize}

\begin{sloppypar}
\item {\em targeted expansion}: the user can avoid creation of non-clique subgraphs by making $expandable(s, \delta)$ return \texttt{true} only when vertex $\delta$ is in $P_s$ (i.e., adding $\delta$ and its relevant edges to $s$ surely leads to a clique).
In this case, every subgraph $s$ obtained through expansion is a clique (i.e., matches the user's interest) and therefore $relevant(s)$ needs to always return \texttt{true}.
\end{sloppypar}

\item {\em prioritized expansion}: the user can implement $priority(s)$ so that it returns $(|V_s|, |P_s|)$ where $V_s$ is the set of vertices in $s$ and $P_s$ is the set of vertices that may be added to $s$ while forming a larger clique. When lexicographic ordering is applied to such priority values, our framework expands larger cliques before smaller cliques and, for cliques of the same size, expands a more promising clique (i.e., clique that is likely to expand into larger cliques) before others. 

\item {\em pruning}: the user can enable pruning by instrumenting $dominated(s, s')$ to return \texttt{true} if $|V_s| + |P_s|$ (the maximum possible size of the cliques into which $s$ can expand) is smaller than $|V_{s'}|$ (the size of clique $s'$).\footnote{This pruning condition was first introduced by Carraghan et al.~\cite{carraghan.orl1990.clique}. The original work by Carraghan et al., however, does not specify any criterion for prioritizing subgraphs.}

\end{itemize}

Figure~\ref{fig:cliques_prioritized} illustrates how our framework can efficiently find the maximum clique ($s_9$ in Figure~\ref{fig:cliques_process}) given the above custom functions and the data graph from Figure~\ref{fig:cliques_graph}. 
Our framework first creates unit cliques $s_1$, $s_2$, $s_3$, and $s_4$ while assigning priorities to them (1).
It then dequeues $s_1$ (i.e., the clique with the highest priority), adds $s_1$ to the result set, expands $s_1$ into $s_5$ and $s_6$, and then enqueues $s_5$ and $s_6$ (2).
Next, it dequeues $s_5$, adds $s_5$ to the result set, removes $s_1$ from the  result set, and expands $s_5$ into $s_9$ (3).
Then, it dequeues $s_9$, and replaces $s_5$ in the result set with $s_9$ (4).
At this point, $s_2$, $s_3$, $s_4$, and $s_6$ can be pruned out since they cannot expand into cliques as large as $s_9$.

\noindent
{\bf Discussion.}
To the best of our knowledge, our subgraph discovery framework is the first one that supports {\em prioritized subgraph expansion}, an ability to expand more {\em promising} subgraphs (i.e., subgraphs whose expansions are more likely to quickly fill the result set with high-priority subgraphs) before others, thereby facilitating {\em early} and {\em effective} pruning.
The benefits of the \oursyst over prior subgraph discovery systems~\cite{teixeira.sosp15.arabesque, quamar2014nscale, chen2018g} which lack prioritized expansion are evident in Figures~\ref{fig:cliques_process} and \ref{fig:cliques_prioritized}.
For example, Arabesque~\cite{teixeira.sosp15.arabesque} must expand all smaller subgraphs (e.g., all subgraphs of size 1) before any larger one (e.g., $s_5$), inherently limiting pruning opportunities (as opposed to ours that can expand $s_5$ before $s_2$, $s_3$, $s_4$, and $s_6$ and then prune out the latter subgraphs).
Arabesque also has to create all subgraphs ({\em exhaustive expansion}) and then filter out irrelevant ones such as non-cliques $s_{10}$, $s_{11}$ and $s_{12}$ ({\em post-filtering}).
On the other hand, our framework can prevent creation of such irrelevant subgraphs through {\em targeted expansion}.

\subsection{Aggregate Computation}
\label{subsec:aggregate}

\begin{algorithm}[]
\small
\SetKw{And}{and}
\SetKw{Or}{or}

create a new aggregator $\agg$;\\
\For {$\textnormal{vertex}~v: V$ \textnormal{or} \textnormal{edge}~$e: E$} {
    create a subgraph $s$ containing only vertex $v$ or edge $e$;\\
    $g \leftarrow key(s)$; {// grouping key (e.g., pattern) from $s$}\\
    $S \leftarrow \agg[g]$; {// the subgraph group associated with $g$ in $\agg$}\\
    add $s$ to $\mathcal{S}$;\\
}
\For {each subgraph group $\mathcal{S}$ in $\agg$} {
    insert $\mathcal{S}$ into  $Q$;
} 
\While {$|Q| > 0$} {
    $\mathcal{S} \leftarrow remove\_max(Q)$;\\
    \If{$relevant(\mathcal{S})$~\And\\ $~~~(|R| < k~\Or~priority(\mathcal{S}) \ge priority( k\textnormal{-}th(R)))$} {
        insert $\mathcal{S}$ into result set $R$;\\
        \If{$|R| > k$}{
            remove from $R$ each entry $\epsilon$ such that $priority(\epsilon) < priority(k\textnormal{-}th(R))$;\\     
        }
    }
    \If{$|R| < k~\Or~!dominated(\mathcal{S}, k\textnormal{-th}(R))$} {
        create a new aggregator $\agg'$;\\
        \For{each subgraph $s$ in $\mathcal{S}$} {
            \For {each neighboring vertex (or edge) $\delta$ of  $s$} {
                \If{$expandable(s,\delta)$} { 
                    create subgraph $s'$ by adding $\delta$ to $s$;\\
                    $g' \leftarrow key(s')$;\\
                    $\mathcal{S}' \leftarrow \agg'[g']$;\\
                    add $s'$ to $\mathcal{S}'$;\\  
                }            
            }
        }
        \For {each subgraph group $\mathcal{S}'$ in $\agg'$} {
            \If {$|R| < k~\Or~!dominated(\mathcal{S}', k\textnormal{-th}(R))$} {
                insert $\mathcal{S}'$ into $Q$;\\
            }
        } 
    }
}
\caption{Aggregate Computation}\label{alg:whole_process_agg}
\end{algorithm}

Some subgraph discovery computations require grouping of subgraphs and aggregation of certain properties. 
Consider the case of finding the $k$ most frequent patterns having two edges in the graph from Figure~\ref{fig:patterns_graph}. 
As illustrated in Figure~\ref{fig:patterns_process}, this computation requires the ability to {\em group} subgraphs according to a certain feature (e.g., pattern) and then obtain an {\em aggregate} value (e.g., frequency) from all of the subgraphs within each group.
Algorithm~\ref{alg:whole_process_agg} shows our extension to the basic computational framework (Algorithm~\ref{alg:whole_process_noAgg}) for the above case as well as other {\em aggregate} subgraph discovery computations.

The main differences of our aggregate computation framework (Algorithm~\ref{alg:whole_process_agg}) compared to the basic (non-aggregate) framework (Algorithm~\ref{alg:whole_process_noAgg}) are as follows:
\begin{enumerate}

\item A new user-specified function, $key(s)$, returns the {\em grouping key} (e.g., pattern) of subgraph $s$. 
Also, an aggregator $\mathcal{G}$ associates each grouping key (e.g., pattern) with the group of subgraphs having that grouping key.
For a grouping key $g$, $\mathcal{G}[g]$ denotes the {\em subgraph group for $g$} (the group of subgraphs whose grouping key is $g$).

\item Analogous to the basic (non-aggregate) framework, prioritization, insertion into the result set, and pruning of {\em subgraph groups} are specified by  functions $priority(S)$, $relevant(S)$, and $dominated(S, S')$ where $S$ and $S'$ are {\em subgraph groups} (as opposed to {\em subgraphs} in the basic framework).
As in the basic framework, $expandable(s, \delta)$ is applied to a subgraph $s$ and a vertex (or edge) $\delta$.

\end{enumerate}

The steps of our aggregate computation framework (Algorithm~\ref{alg:whole_process_agg}) are as follows: 
It first constructs a new aggregator $\mathcal{G}$ (line 1) and creates, for each vertex/edge (line 2), a subgraph $s$ containing that vertex/edge (line 3), finds the grouping key $g$ of $s$ (line 4), and inserts $s$ into the subgraph group for $g$ (lines 5-6).
It then inserts each subgraph group into a priority queue $Q$ (lines 7-8).
Next, as long as $Q$ contains subgraph groups (line 9), it repeatedly dequeues and processes the subgraph group $S$ with the highest priority (lines 10-27).
If $S$ matches the user's interest (line 11) and if the result set $R$ is not yet full or $S$ has as high a priority as the $k$-th entry in $R$ (line 12), it adds $S$ to $R$ (line 13) and removes unnecessary entries from $R$ (lines 14 and 15).
It also examines if subgraph group $S$ can be safely ignored (i.e., unnecessary  to expand subgraphs in $S$) (line 16).
If not, for each subgraph $s$ in $S$ (line 18), it considers each neighboring vertex/edge $\delta$ of $s$ (line 19).
If adding $\delta$ to $s$ is adequate (line 20; e.g., $s$ has fewer than 2 edges), it expands $s$ into $s'$ by adding $\delta$ (line 21), finds the grouping key $g'$ of $s'$ (line 22), and then inserts $s'$ into the subgraph group for $g'$ (lines 23 and 24).
After grouping new subgraphs as above, for each subgraph group $S'$ (line 25), it examines if $S'$ can be ignored (line 26).
If not, it inserts $S'$ into $Q$ (line 27).

\noindent
{\bf Example: Top-$\boldsymbol{k}$ Frequent Pattern Mining.}
Consider the problem of finding the most frequent 2-edge patterns.
A user can efficiently solve this problem by implementing custom functions as follows (for the implementation of these functions, refer to Section~\ref{subsec:api_agg}):
\begin{itemize}

\item {\em subgraph expansion}: to obtain all subgraphs consisting of up to two edges, the $expandable(s, \delta)$ function needs to return \texttt{true} if $s$ contains less than two edges.
To include 2-edge patterns in the result set, the $relevant(S)$ function needs to return \texttt{true} if the grouping key (i.e., the pattern) of $S$ has two edges.

\item {\em prioritized expansion}: the user can implement $priority(S)$ so that it returns $(m(S), f(S))$ where $m(S)$ denotes the number of edges in the pattern associated with subgraph group $S$ and $f(S)$ denotes the frequency of that pattern. 
If lexicographic ordering is applied to such priority values, our framework processes larger patterns before smaller patterns, and, for patterns of the same size, processes more promising (i.e., frequent) patterns before others.

\item {\em pruning}: the user can enable pruning by (i) using a pattern frequency metric with anti-monotonicity (e.g., {\em minimum image-based support}~\cite{bringmann.pakdd08.minimImageBased}), which guarantees that any super-pattern $p'$ of $p$ cannot have a higher frequency than $p$ (i.e., $f(p') \le f(p)$), and (ii) by making $dominated(S, S')$ return \texttt{true} if $f(S) < f(S')$.
When $S'$ is the $k$-th entry in the result set and $dominated(S, S')$ returns \texttt{true}, all subgraphs in $S$ can be safely ignored since expansions of them cannot affect the result set (due to anti-monotonicity, all subgraph patterns obtained through these expansions would have a lower frequency than the pattern associated with $S'$).

\end{itemize}

\begin{figure}[t]
    \includegraphics[width=.48\textwidth, trim={0 .6cm 0 0}]{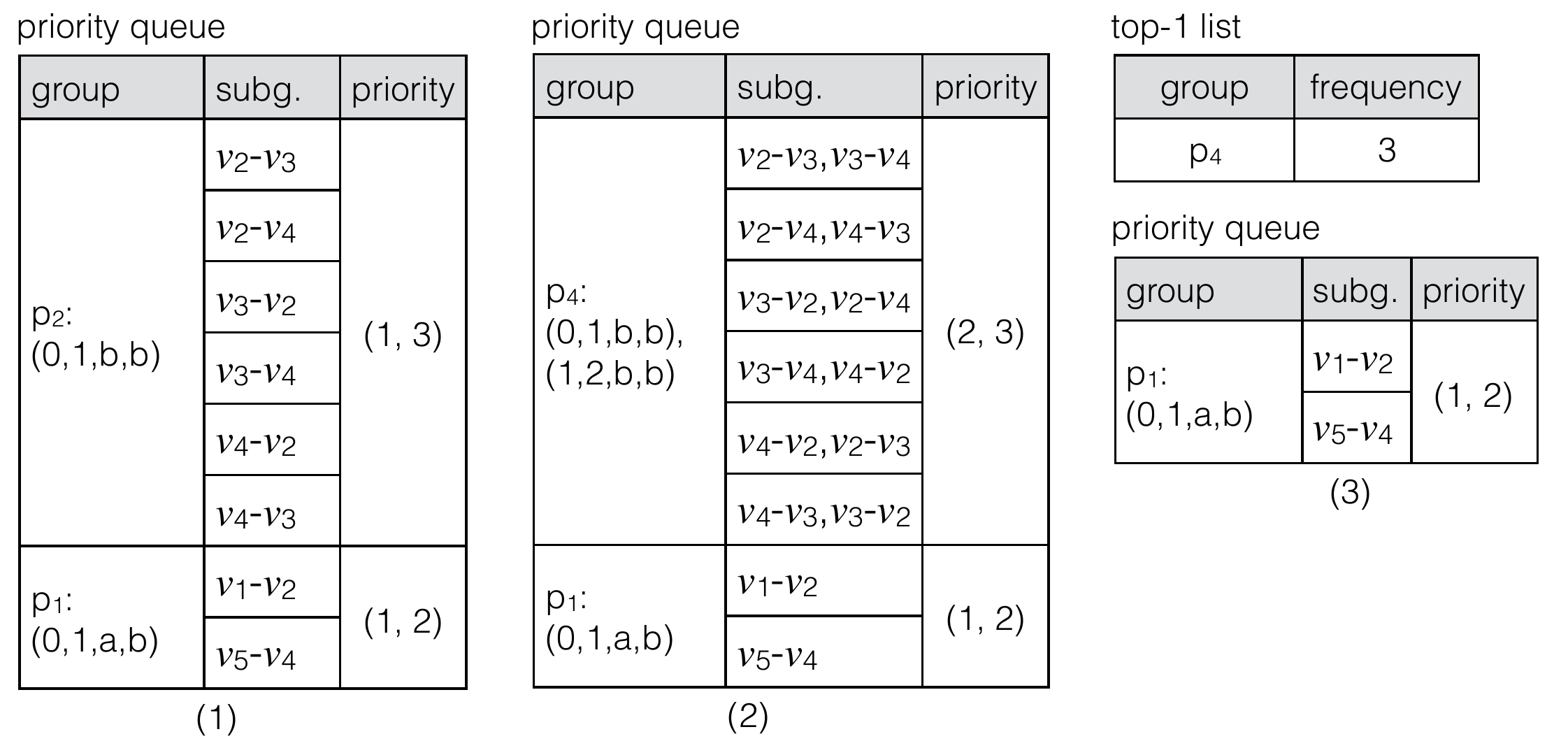}
    \caption{Most Frequent Pattern Mining}\label{fig:patterns_prioritized}
\end{figure}

\begin{figure}[h]
    \includegraphics[width=.44\textwidth, trim={0 .2cm 0 0}]{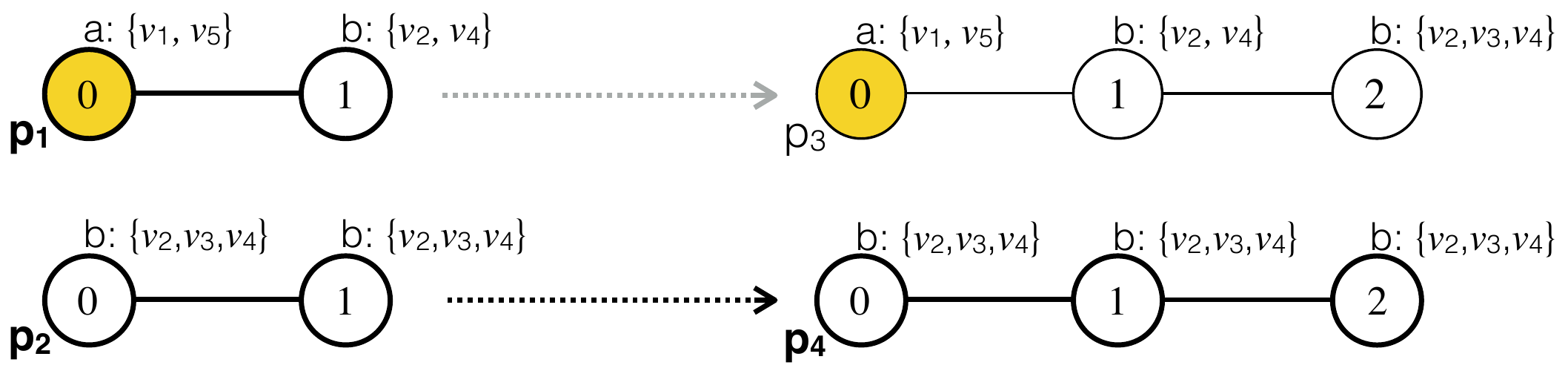}
    \caption{Relationships between Patterns
    }\label{fig:patterns_induced}
\end{figure}

Figure~\ref{fig:patterns_prioritized} illustrates how our framework can efficiently find the most frequent pattern ($p_4$) given the above custom functions and the graph from Figure~\ref{fig:patterns_graph}.
In contrast to the examples shown in Figure~\ref{fig:patterns_process} where each subgraph expansion adds an edge to a subgraph ({\em edge-oriented expansion}) and Figure~\ref{fig:cliques_process} where each subgraph expansion adds, to a subgraph, a vertex and its edges connected to a vertex in that subgraph ({\em vertex-oriented expansion}), this example uses a different subgraph expansion approach which we call {\em pattern-oriented expansion}.

The {\em pattern-oriented expansion} approach creates each subgraph by adding a series of directed edges and expresses each subgraph pattern using the DFS code~\cite{yan.icdm02.gspan}.
The DFS code represents each directed edge as $(i, j, L(i), L(i, j), L(j))$, where $i$ and $j$ are vertex IDs, $L(i)$ and $L(j)$ denote the labels of vertices $i$ and $j$, and $L(i, j)$ denotes the label of the edge from $i$ to $j$ (in our examples, $L(i, j)$ are omitted since edges are not assigned any label).
For example, in Figure~\ref{fig:patterns_prioritized}, it initially creates 8 one-edge subgraphs.
It obtains pattern $p_1$ from a subgraph containing the edge from vertex $v_1$ to vertex $v_2$ (denoted edge  $v_1$-$v_2$) and another subgraph containing edge $v_5$-$v_4$.
It expresses $p_1$ as $(0, 1, a, b)$ since $p_1$ consists of only one edge from vertex $0$ with label $a$ to vertex $1$ with label $b$.

The {\em pattern-oriented expansion} approach constructs a subgraph $s$ only if its DFS code (i.e., the DFS code of its pattern) is {\em minimal}~\cite{yan.icdm02.gspan}.
For example, it constructs a subgraph containing edge $v_1$-$v_2$, but not a subgraph containing edge $v_2$-$v_1$ since the DFS codes of these subgraphs are $(0, 1, a, b)$ and $(0, 1, b, a)$, respectively, and $(0, 1, a, b) < (0, 1, b, a)$.
Similarly, it does not add edge $v_4$-$v_5$ to edge $v_3$-$v_4$ since the resulting DFS code $[(0, 1, b, b),$ $(1, 2, b, a)]$ is not minimal (adding edge $v_4$-$v_3$ to edge $v_5$-$v_4$ leads to a smaller code $[(0, 1, a, b), (1, 2, b, b)]$).
The above subgraph construction condition provides the following guarantee:
\begin{property}\label{property:pattern}
Let pattern $p'$ be a super-pattern of $p$ such that the DFS codes of $p$ and $p'$ are $[c_1, c_2, \cdots, c_{n-1}]$ and $[c_1, c_2, \cdots, c_{n}]$, respectively.
Under pattern-oriented expansion, all of the subgraphs matching pattern $p'$ can be obtained by expanding only the subgraphs matching $p$~\cite{yan.icdm02.gspan}.  
\end{property}


Due to Property~\ref{property:pattern}, the frequency of $p_3$ can be obtained by expanding {\em only} the subgraphs matching $p_1$ (Figure~\ref{fig:patterns_induced}).
On the other hand, in Figure~\ref{fig:patterns_process} where edge-oriented expansion is applied, the frequency of $p_3$ can be calculated only after expanding both the subgraphs matching $p_1$ and the subgraphs matching $p_2$.

In Figure~\ref{fig:patterns_prioritized}, our framework first forms two sugraph groups from 8 one-edge subgraphs (1).
Between these groups, it selects the group for pattern $p_2$ (i.e., the group with the highest priority), expands the subgraphs in that group, and obtains a new group of subgraphs matching pattern $p_4$ (2).
Since pattern $p_4$ has two edges (i.e., matches the user's interest), it inserts the subgraph group for pattern $p_4$ in the result set (3).
At this point, it can ignore the subgraphs matching pattern $p_1$ since the patterns obtained by expanding these subgraphs cannot be as frequent as $p_4$.

\noindent
{\bf Discussion.}
To the best of our knowledge, our work mentioned above is the first top-$k$ aggregate subgraph discovery framework which supports both {\em prioritization} and {\em pruning} which can drastically improve performance (Section~\ref{subsec:result_patterns}).
Among the existing subgraph discovery systems~\cite{teixeira.sosp15.arabesque, quamar2014nscale, chen2018g}, NScale~\cite{quamar2014nscale} and GMiner~\cite{chen2018g} do not support aggregate computations.
Arabesque~\cite{teixeira.sosp15.arabesque} adopts {\em edge-oriented} subgraph expansion and thus has to expand {\em all smaller} subgraphs before any larger subgraphs (Figure~\ref{fig:patterns_process}), thereby inevitably limiting prioritization and pruning opportunities.
On the other hand, due to its use of {\em pattern-oriented expansion}, our framework can quickly find each frequent pattern (e.g., $p_4$ in Figures~\ref{fig:patterns_prioritized} and \ref{fig:patterns_induced}) after expanding {\em only one} group of subgraphs matching a sub-pattern (e.g., $p_2$) and thus can enable {\em early} and {\em effective} pruning (e.g., pruning of subgraphs matching $p_1$; no creation of subgraphs matching $p_3$).

\section{API}
\label{sec:api}


In this section, we present details of our API along with actual implementation code for clique discovery (Section~\ref{subsec:api_noAgg}), pattern mining (Section~\ref{subsec:api_agg}), and subgraph isomorphism (Section~\ref{subsec:api_indexing}).


\subsection{API for Non-Aggregate Computation}
\label{subsec:api_noAgg}

A user who wants to implement a non-aggregate subgraph discovery computation (Section~\ref{subsec:non-aggregate}) needs to choose a subgraph expansion approach between vertex-oriented expansion (Figure~\ref{fig:cliques_process}) and  edge-oriented expansion (Figure~\ref{fig:patterns_process}).
Then, the user needs to create a new class, in Java, that extends either the \texttt{VertexOrientedSubgraph} type with the \texttt{expandable(Vertex v)} method or the \texttt{EdgeOrientedSubgraph} type with \texttt{expandable(Edge e)} depending on the chosen expansion approach.
Implementing {\em only} the \texttt{relevant()} method enables the  most preliminary form of subgraph discovery (without result ranking, pruning, and prioritization).
The user can support targeted expansion by implementing \texttt{expandable(Vertex v)} or \texttt{expandable(Edge e)}, result ranking and prioritization by implementing \texttt{priority()}, and pruning by implementing  \texttt{dominated(S o)}, where \texttt{S} is a Java generic type that refers to the class being created.
These methods correspond to the $expandable(s, \delta)$, $relevant(s)$, $priority(s)$, and $dominated(s, s')$ functions in Table~\ref{table:functions}.

\begin{figure}[h!]
\hspace{.3cm}                       
\begin{lstlisting}[caption=Clique Discovery (\texttt{SubgraphCD}),label=listing:cliques]
public HashSet<Vertex> p = null;

HashSet<Vertex> p() {
	if (p == null) {
		p = new HashSet<Vertex>();
		for (Vertex v : vertices())
			if (v.equals(seed()))
				p.addAll(neighbors(v));
			else {
				p.remove(v);
				p.retainAll(neighbors(v));
			}
	}
	return p;
}
\end{lstlisting}
\end{figure}

\noindent
{\bf Example: Maximum Clique Discovery.}
As explained in Section 3.2, the maximum clique discovery computation can be implemented as follows (using a method \texttt{p()} that returns the set of  vertices that can be added to a clique while forming a larger clique): 
(i) \texttt{expandable(Vertex v)} returns \texttt{p().contains(v)}; 
(ii) \texttt{relevant()} returns \texttt{true}; 
(iii) \texttt{dominated(S o)} returns \texttt{(vertexCount() + p().size() < o.vertexCount())}; (iv) \texttt{priority()} returns \texttt{new double[] \{ vertexCount(), p().size() \}}.
Listing~\ref{listing:cliques} shows a code snippet that implements the method \texttt{p()} mentioned above based on the CP algorithm~\cite{carraghan.orl1990.clique}.
For the first vertex in the current clique (line 7), it adds to a set \texttt{p} the neighbors of that vertex (i.e., the vertices that can be added to the clique while forming a 2-vertex clique) (line 8).
For every vertex \texttt{v} in the current clique except the first vertex, \texttt{p} needs to exclude \texttt{v} (line 10; since \texttt{v} is already in that clique) and retain only the vertices connected to \texttt{v} (line 11; i.e., vertices that can belong to a clique containing \texttt{v}).

\subsection{API for Aggregate Computation}
\label{subsec:api_agg}

Implementation of an aggregate subgraph discovery computation (Section~\ref{subsec:aggregate}) requires creation of two classes, one extending the \texttt{SubgraphGroup} type which contains the \texttt{key(Subgraph s)}, \texttt{relevant()}, \texttt{dominated(S o)}, \texttt{priority()} methods (corresponding to the custom functions $key(s)$, $relevant(S)$, $dominated(S, S')$, and $priority(S)$ in Section~\ref{subsec:aggregate}) and another class for representing subgraphs.
To enable pattern-oriented expansion (Section~\ref{subsec:aggregate}), the latter class must extend the \texttt{PatternOrientedSubgraph} type which includes the \texttt{expandable(Edge e)} method.

\begin{figure}[h!]
\hspace{.3cm}                       
\begin{lstlisting}[caption=Pattern Mining (\texttt{SubgraphGroupPM}),label=listing:patterns]
Integer f = null;

int f() {
	if (f == null) {
		MinimumImageSupport support = new MinimumImageSupport();
		for (Subgraph s : subgraphs())
			support.aggreagate(s);
		f = support.frequency();
	}
	return f;
}
\end{lstlisting}
\end{figure}

\noindent
{\bf Example: Top-$\boldsymbol{k}$ Frequent Pattern Mining.}
As explained in  Section~\ref{subsec:aggregate}, for  top-$k$ frequent pattern mining, the class implementing \texttt{PatternOrientedSubgraph} needs to include the \texttt{expandable(Edge e)} method which returns \texttt{(edgeCount() < M)}.
Furthermore, the class implementing the \texttt{SubgraphGroup} type needs to include the following methods (as shown in Listing~\ref{listing:patterns}, method \texttt{f()} returns the frequency of the pattern associated with the current subgraph group):
(i) \texttt{key(Subgraph s)} returns \texttt{pattern(s)}; 
(ii) \texttt{relevant()} returns \texttt{(((Pattern) key()).edgeCount() == M)};
(iii) \texttt{dominated(S o)} returns \texttt{(f() < o.f())}; and
(iv) \texttt{priority()} returns \texttt{new double[] \{ ((Pattern) key()).edgeCount(), f() \}}.

\subsection{API for Indexing}
\label{subsec:api_indexing}

Indexing techniques for subgraph discovery typically add index entries for each vertex~\cite{gupta.icde14.topksubgraph, zou2007top}.
To implement such techniques, users need to implement a class extending the \texttt{Vertex} type with the following methods: (i) \texttt{put(k1, k2, $\cdots$, kn, v)} which associates a key comprising \texttt{k1}, \texttt{k2}, $\cdots$, and \texttt{kn} with a value \texttt{v} as an attribute of the vertex, (ii) \texttt{get(k1, k2, $\cdots$, kn)} which returns the value associated with the key comprising \texttt{k1}, \texttt{k2}, $\cdots$, and \texttt{kn}, and (iii) \texttt{initialize()} which is invoked automatically for every vertex when the data graph is loaded into the system. 

\begin{figure}[t]
\hspace{.3cm}                       
\begin{lstlisting}[caption=Subgraph Isomorphism Search (\texttt{VertexIS}),label=listing:iso]
public void initialize() {
	for (int i = 1; i <= H; i++)
		for (Vertex n : neighbors(i)) {
			Integer s = (Integer) get(n.label(), i);
			put(n.label(), i, s == null ? n.get("score") : 
				Math.max(s, (Integer) n.get("score")));
		}
}
\end{lstlisting}
\end{figure}

\begin{figure}[h!]
    \includegraphics[width=.5\textwidth, trim={0 0cm 0 0}]{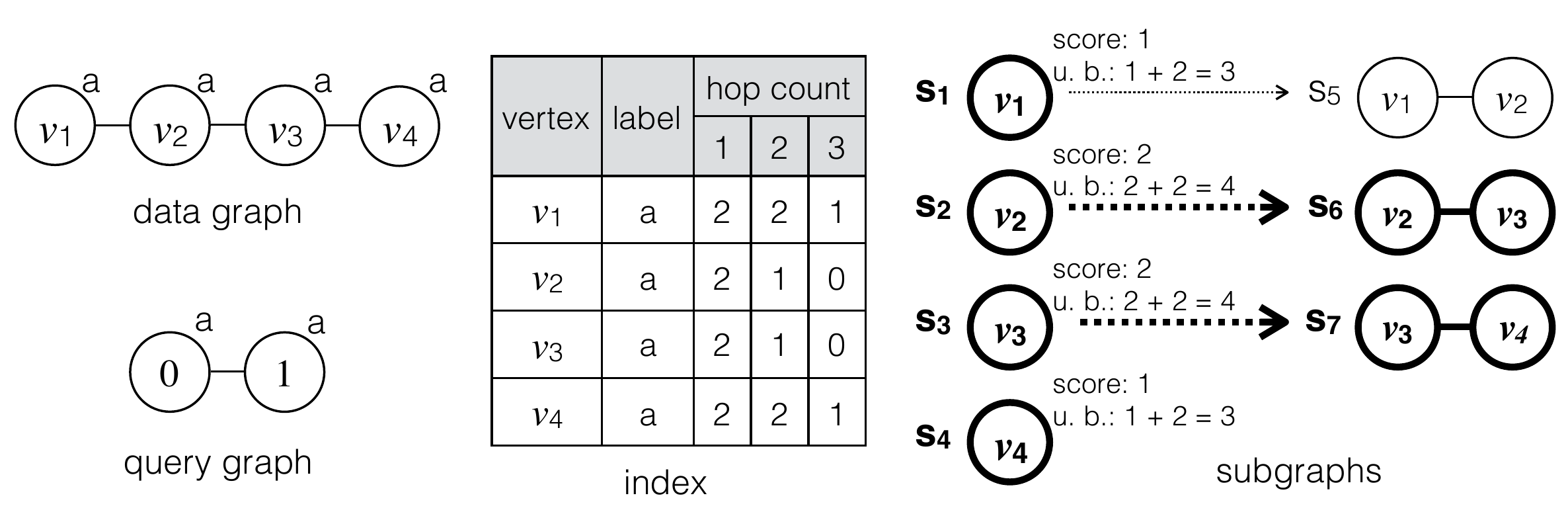}\caption{Subgraph Isomorphism Example }\label{fig:iso_example}
\end{figure}

\noindent
{\bf Example: Top-$\boldsymbol{k}$ Subgraph Isomorphism.}
Top-$k$ Subgraph Isomorphism (SI) discovery~\cite{gupta.icde14.topksubgraph} aims to find,  in the input graph,  the $k$ highest-scored subgraphs which are isomorphic to a query graph.
In this example, we define the score of each subgraph as the sum of the degree (e.g., the number of citations of each paper in a citation network) of the vertices in that subgraph.
We enable targeted expansion by implementing the \texttt{expandable(Edge e)} method so that it returns \texttt{true} when the addition of \texttt{e} to the current subgraph leads to a larger subgraph that matches a part of the query graph (i.e., is isomorphic to a subgraph of the query graph) based on Ullman's subgraph isomorphism algorithm~\cite{ullmann.j1976.iso}.

To facilitate pruning in \oursyst, we build an index for every vertex in the input graph in a way similar to the work by Gupta et al.~\cite{gupta.icde14.topksubgraph}. 
The index stores, for each hop count $d$ and vertex label $l$, the maximum degree over all vertices that have label $l$ and that are $d$-hop away from the vertex under consideration (Figure~\ref{fig:iso_example} and Listing~\ref{listing:iso}). 
Every subgraph also maintains a look-up map of the vertices in the query graph that are not yet matched to the vertices in that subgraph.
If a subgraph is obtained by repeatedly expanding a subgraph containing a seed vertex $s$, it is possible to derive an upper bound on the scores of the subgraphs that the subgraph can expand into.
This upper bound is defined as the sum of (1) the current score of the subgraph and (2) an upper bound on the score that can be obtained from the un-matched vertices in the look-up map (i.e., $\sum_{v \in \mathcal{M}}{index(s, label(v)}, hop(v))$, where $\mathcal{M}$ denotes the look-up map, $hop(v)$ denotes the distance of $v$ in the query graph from the vertex that corresponds to vertex $s$, $label(v)$ denotes the label of vertex $v$, and $index(s, l, h)$ denotes the value in the index for vertex $s$, label $l$, and hop count $h$.

Figure~\ref{fig:iso_example} shows how the score upper bound explained above can be calculated for each subgraph.
For example, subgraph $s_{2}$  contains vertex $v_2$ which is matched with vertex $0$ in the query graph.
Therefore, its look-up map contains vertex $1$ (which is not yet matched) in the query graph.
The current score of $s_{2}$ is 2 since the degree of its only vertex $v_2$ is 2.
Vertex $1$ in the look-up map is 1 hop away from vertex 0 and its label is $a$.
Since the maximum degree over all vertices in the data graph that has label $a$ and is 1 hop away from vertex $v_2$ is 2 (see the index in Figure~\ref{fig:iso_example}), without expanding $s_{2}$ to include vertex $v_3$, it can be found that an additional score of 2 can be obtained from vertex $1$ in the look-up map.
Therefore, the score upper bound for $s_{2}$ is calculated as 4.
Similarly, the score upper bound for $s_{1}$, $s_{3}$, and $s_{4}$ are calculated as 3, 4, and 3, respectively.
Furthermore, subgraph $s_{2}$ has a higher priority (i.e., upper bound) than $s_{1}$, and thus $s_{2}$ is expanded earlier than $s_{1}$.
In the case of top-1 subgraph isomorphism discovery, $s_{1}$ can be safely discarded after $s_{2}$ is expanded into $s_{6}$ whose final score is 4.
The reason is that any subgraph expanded from $s_{1}$ can only have a lower score than $s_{6}$.
To support pruning and prioritization as explained above, \texttt{dominated(S o)} and \texttt{priority()} need to return \texttt{(score() + u() < o.score())} and \texttt{new double[] \{ edgeCount(), score() + u() \}}, where \texttt{score()} and \texttt{u()} return the score and the score upper bound of the current subgraph, respectively.


\section{System Architecture}
\label{sec:architecture}

\begin{figure}[t]
\center
    \includegraphics[width=.42\textwidth, trim={0 1cm 0 0}]{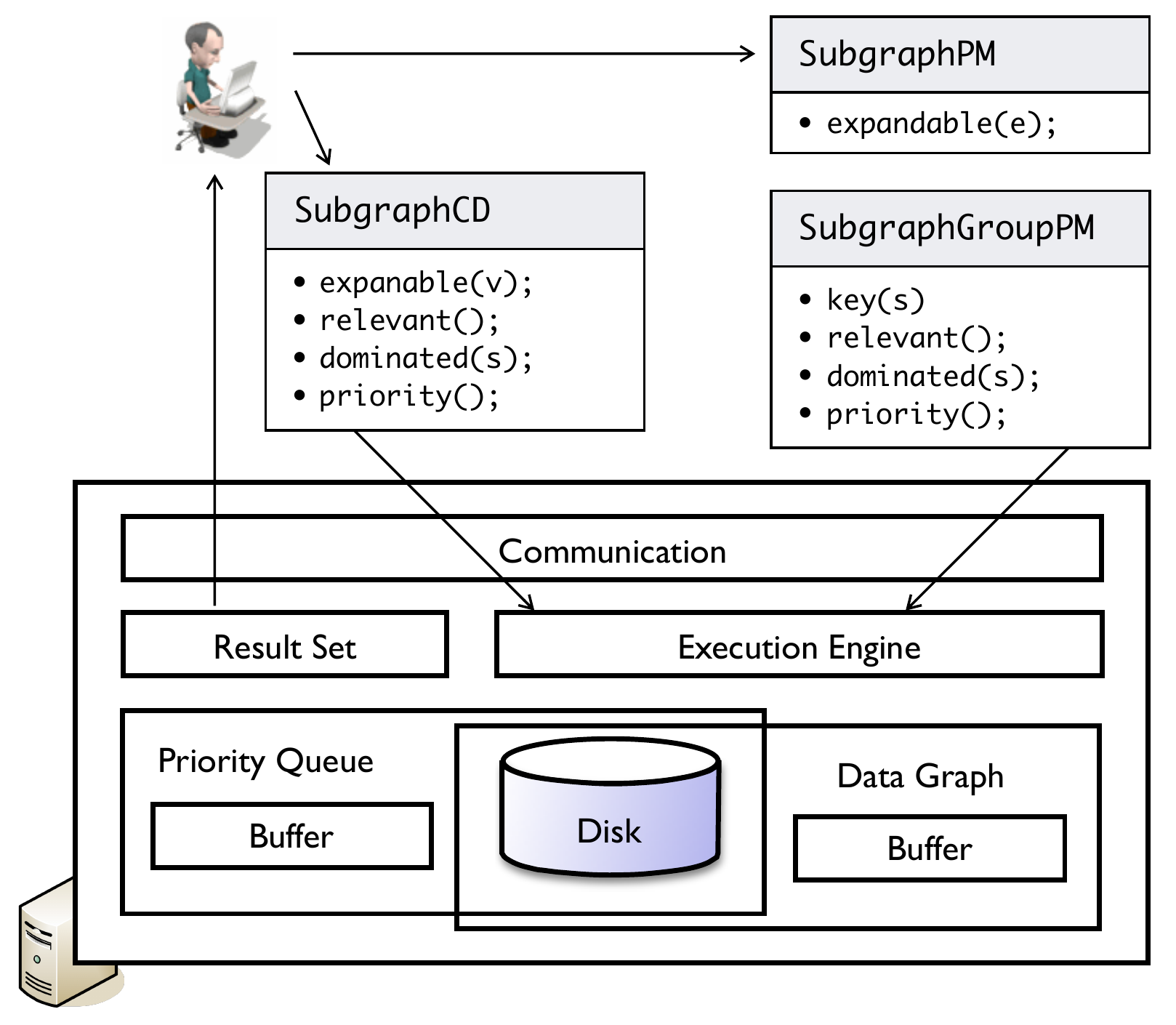}\caption{System Architecture}\label{fig:architecture}
\end{figure}

\begin{figure*}[t]
\footnotesize

\begin{minipage}[b]{0.33\linewidth}
     \subfloat[subgraphs\label{fig:result_cliques_email_candidates}]{%
       \includegraphics[width=0.48\textwidth, trim={1.6cm 0.4 .6cm 0}]{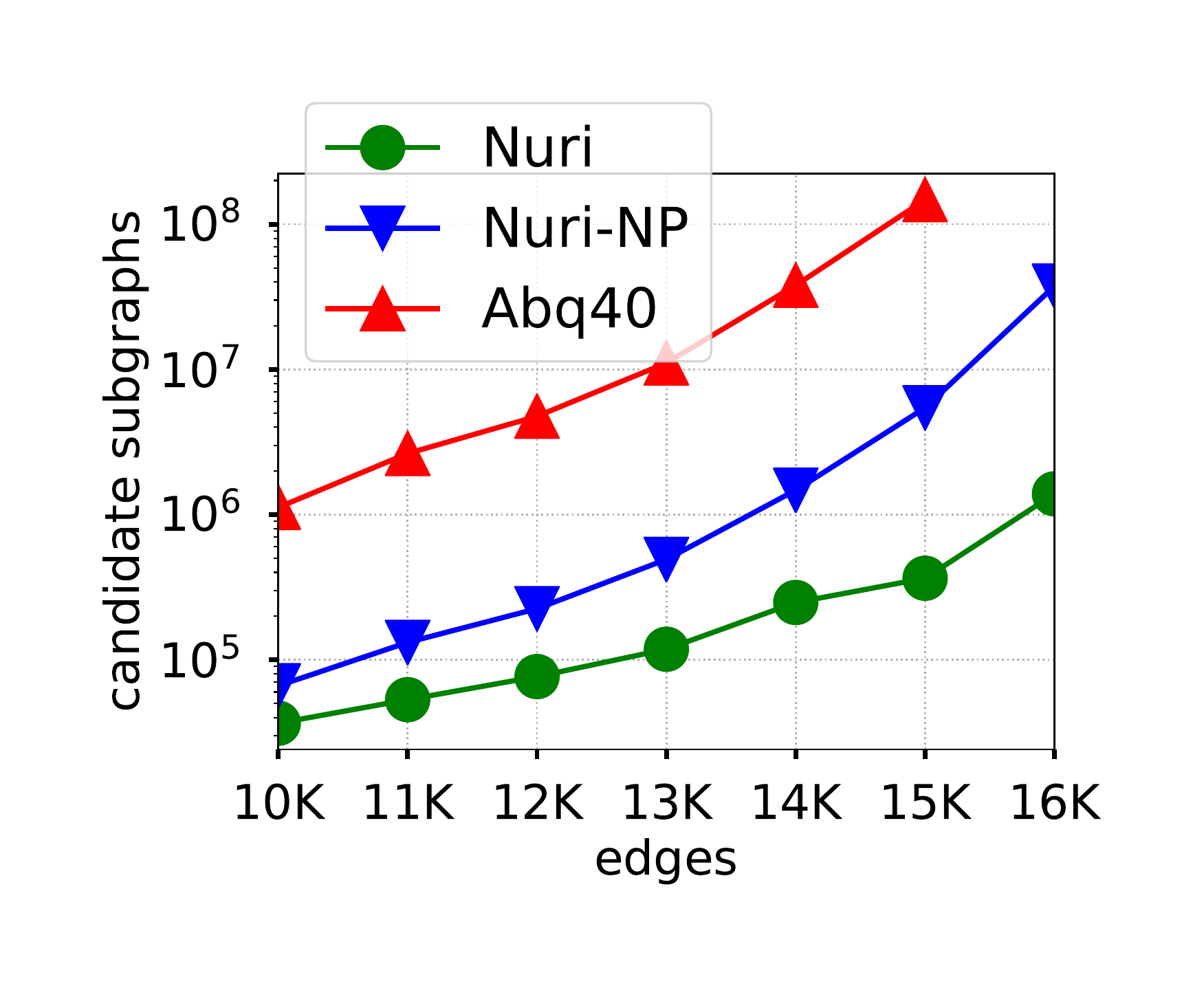}
     }
     \subfloat[time\label{fig:result_cliques_email_time}
     ]{%
       \includegraphics[width=0.48\textwidth, trim={1.6cm 0 .6cm 0}]{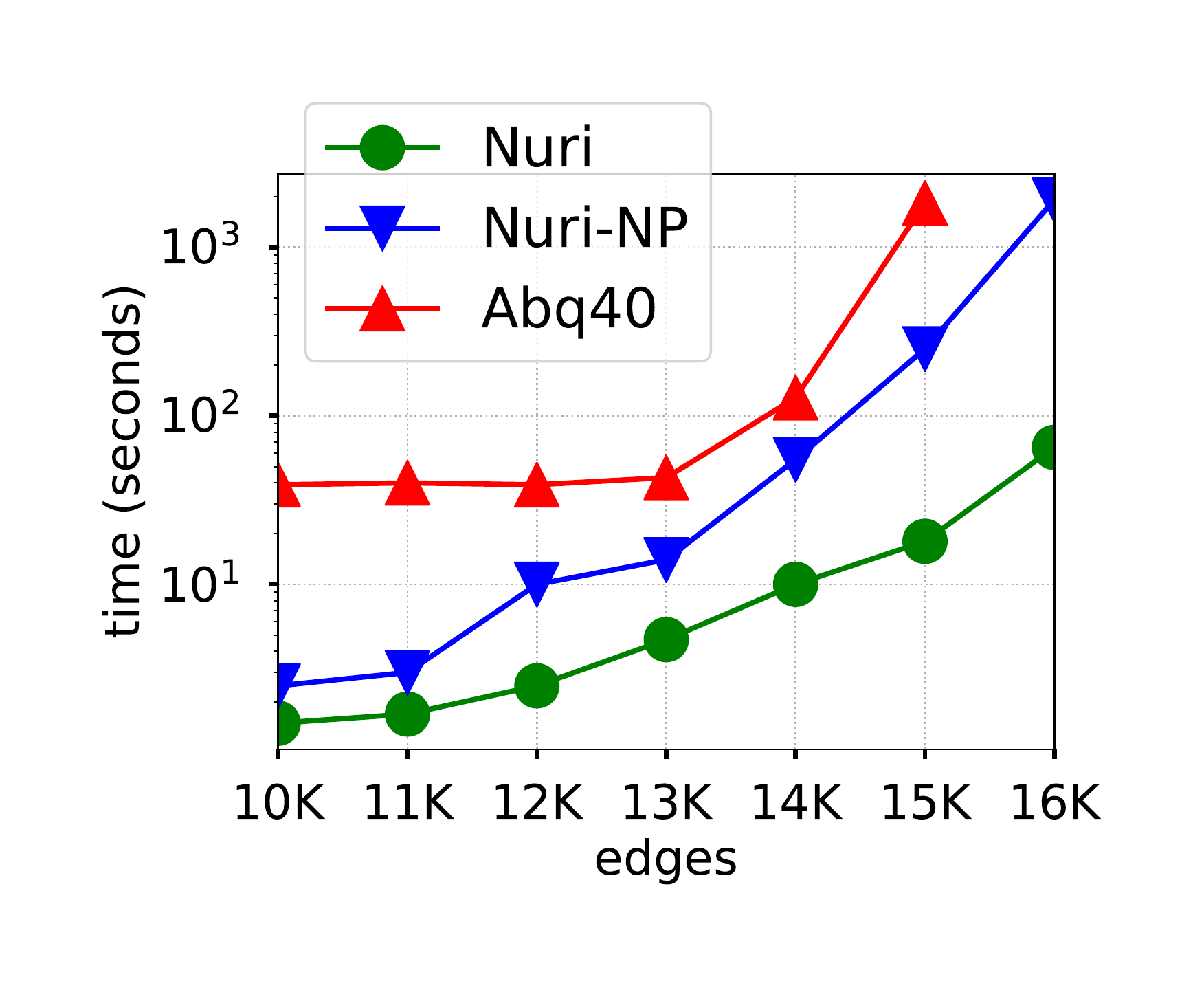}
     }
    \caption{Clique Discovery (\texttt{Email})}\label{fig:result_cliques_email}
\end{minipage}
\begin{minipage}[b]{0.33\linewidth}
     \subfloat[subgraphs\label{fig:result_cliques_MiCo_candidates}]{%
       \includegraphics[width=0.48\textwidth, trim={1.6cm 0 .6cm 0}]{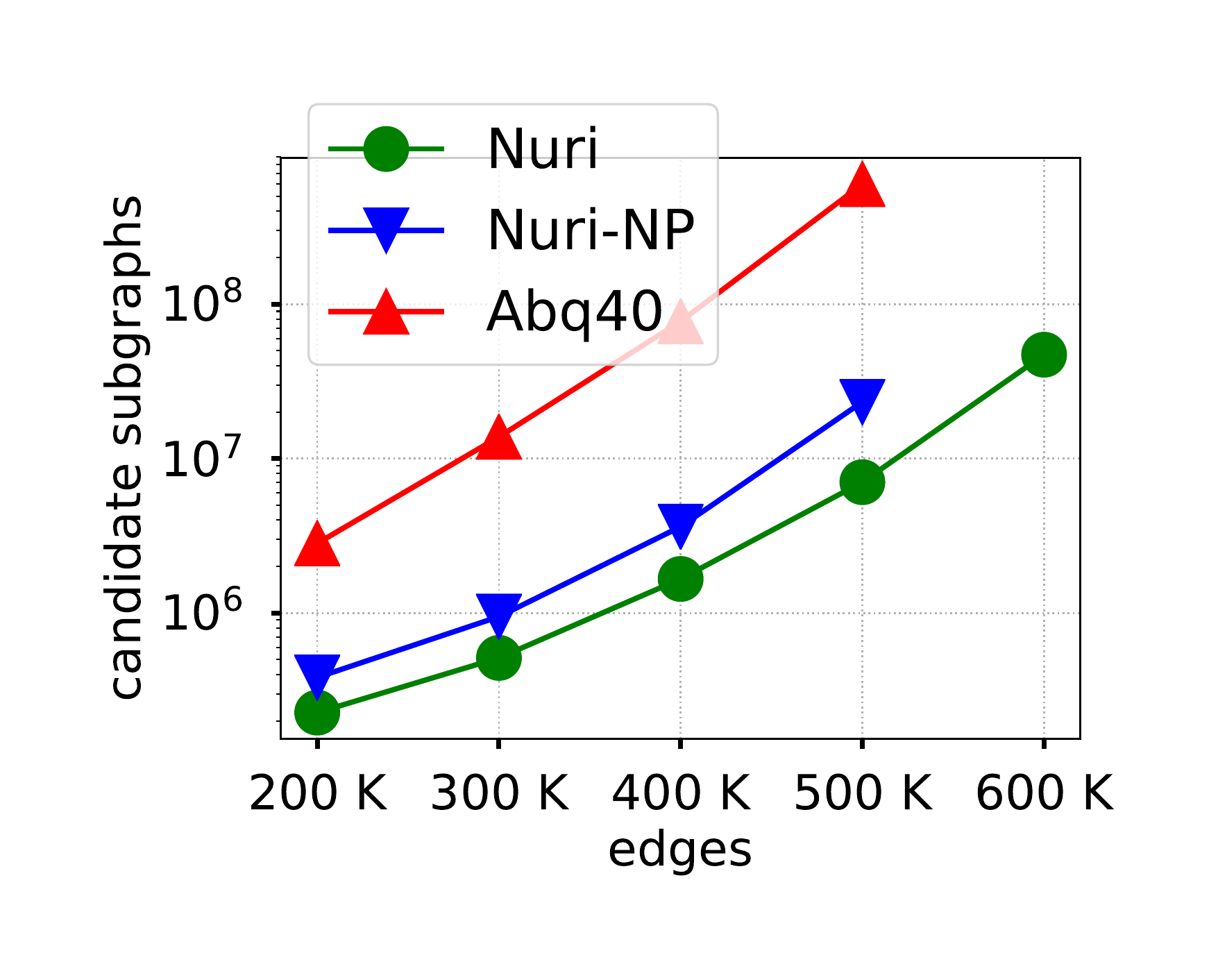}
     }
     \subfloat[time\label{fig:result_cliques_MiCo_time}
     ]{%
       \includegraphics[width=0.48\textwidth, trim={1.6cm 0 .6cm 0}]{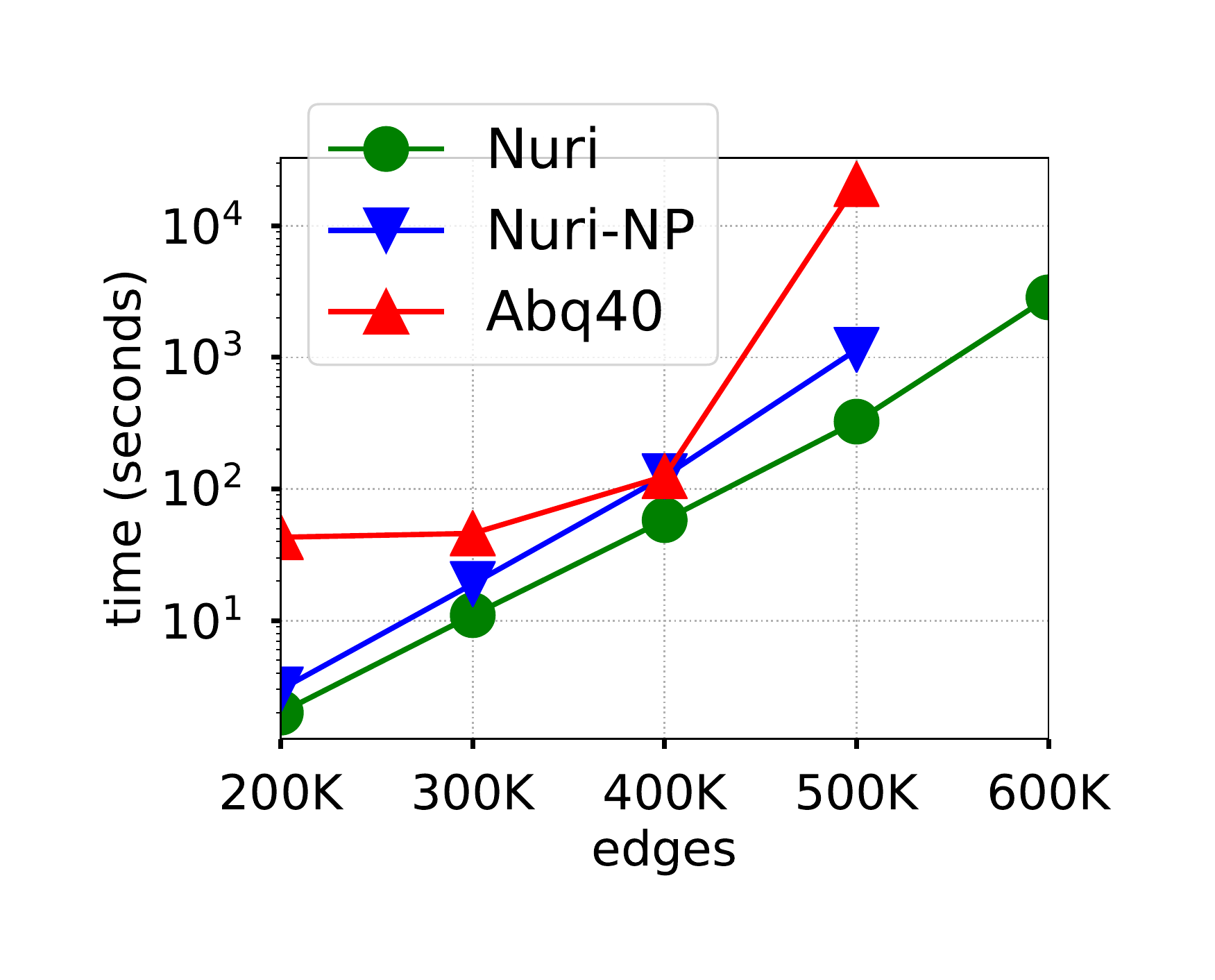}
     }
    \caption{Clique Discovery (\texttt{MiCo})}\label{fig:result_cliques_mico}
\end{minipage}
\begin{minipage}[b]{0.33\linewidth}
     \subfloat[subgraphs\label{fig:result_cliques_youtube_candidates}]{%
       \includegraphics[width=0.48\textwidth, trim={1.6cm 0 .6cm 0}]{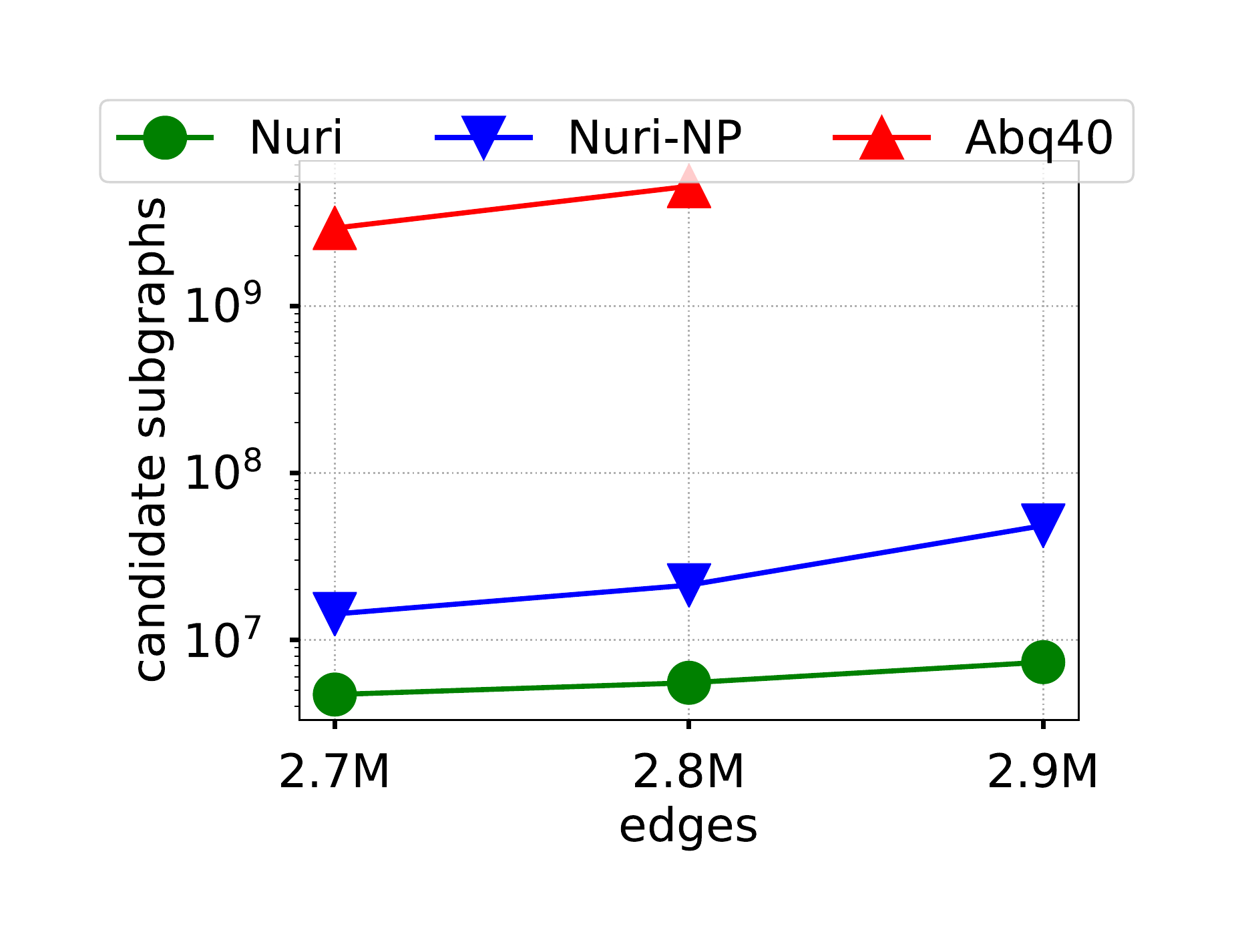}
     }
     \subfloat[time\label{fig:result_cliques_youtube_time}
     ]{%
       \includegraphics[width=0.48\textwidth, trim={1.6cm 0 .6cm 0}]{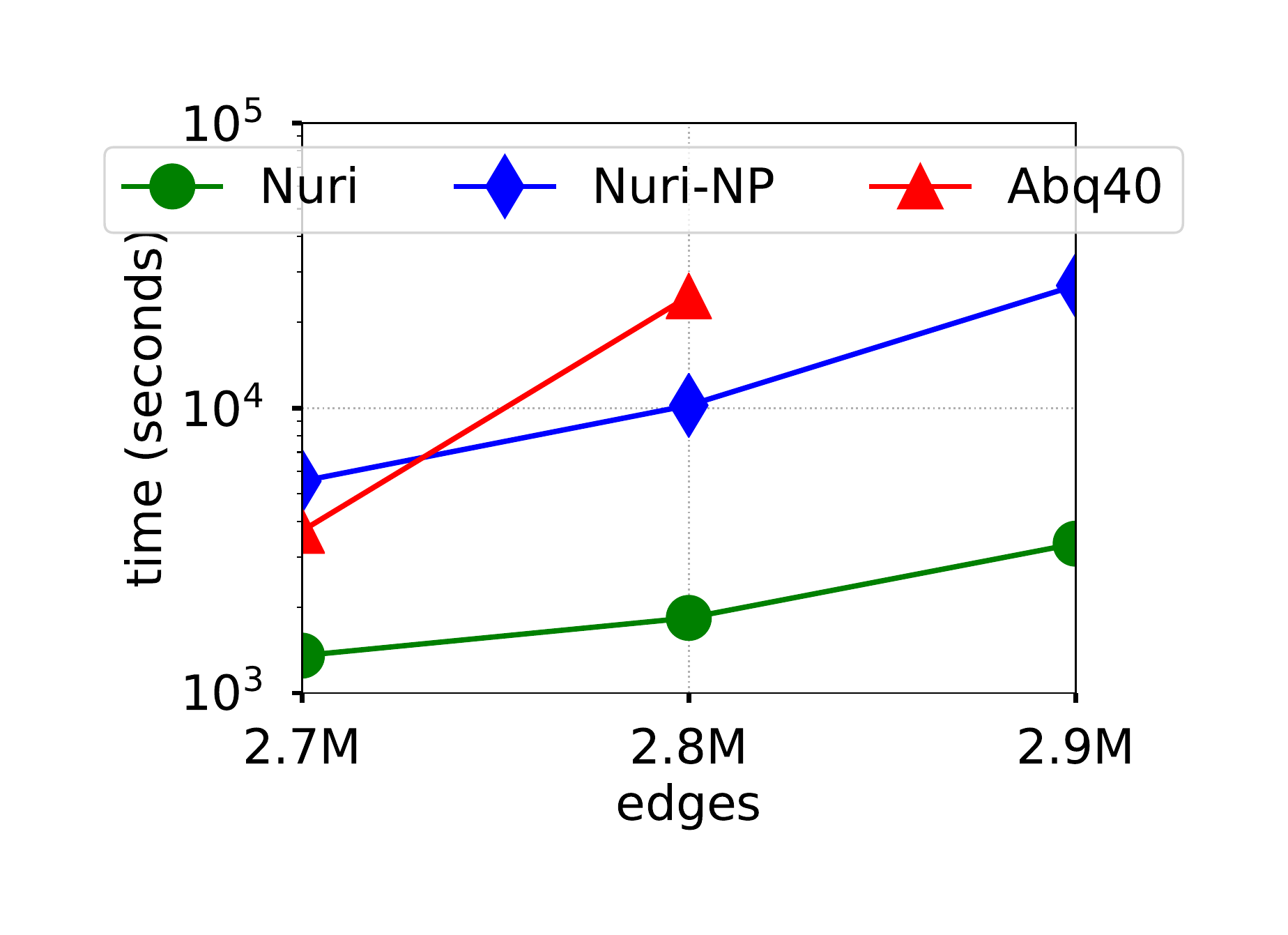}
     }
    \caption{Clique Discovery (\texttt{YouTube})}\label{fig:result_cliques_youtube}
\end{minipage}

\end{figure*}

Figure~\ref{fig:architecture} illustrates the architecture of our system.
When a user submits an implementation of non-aggregate subgraph discovery (e.g., \texttt{SubgraphCD} for clique discovery), the {\em execution engine} carries out the computation by inserting unit subgraphs into the {\em priority queue} and repeats the process of dequeuing the subgraph with the highest priority and expanding it into larger subgraphs while inserting the subgraphs matching the user's interest into the {\em result set} and pruning out irrelevant subgraphs (Algorithm~\ref{alg:whole_process_noAgg}).
When a user submits an implementation of aggregate subgraph discovery (e.g., \texttt{SubgraphPM} and \texttt{SubgraphGroupPM} for pattern mining), the execution engine carries out additional grouping and aggregation operations (Algorithm ~\ref{alg:whole_process_agg}).
For each subgraph discovery computation, the execution engine loads the {\em data graph} from the disk into the main memory.
When each computation completes, the obtained {\em result set} is sent to the user.
The {\em communication} component enables such interactions between the user and the system.

The number of subgraphs kept in the priority queue usually increases exponentially with the size and density of the data graph.
For this reason, we created an implementation of {\em external priority queue} that can store a large number of entries on disk without being limited by the size of the main memory~\cite{fadel1999heaps}.

Our implementation, called {\em virtual priority queue}, has the ability to manage high-priority subgraphs in memory and low-priory subgraphs on disk in a highly efficient manner where the use of disk causes only a slight degradation in the speed of enqueue/dequeue operations (Section~\ref{subsec:result_queue}).
It initially maintains subgraphs using a standard memory-resident priority queue.
Whenever the memory usage becomes higher than a threshold, it retains only half of the subgraphs in the memory-resident queue and stores the others on disk in order of decreasing priority.
The collection of the subgraphs on disk is called a {\em run} (analogous to runs in the context of external sorting~\cite{manker.commun63.sorting, lozinskii.cybernetics68.mergeSort}).
When there are multiple runs (since the memory-resident queue was full multiple times in the past), it can still quickly retrieve subgraphs from all of the runs in order of decreasing priority (analogous to external merge sort~\cite{manker.commun63.sorting, lozinskii.cybernetics68.mergeSort}).
It also applies buffering to read each run with a small number of disk seeks.


\section{Evaluation}
\label{sec:evaluation}

In this section, we explain our setup (Section~\ref{subsec:setup}) for evaluating the effectiveness of \oursyst in comparison to alternatives for clique discovery (CD), pattern mining (PM), and subgraph isomorphism (SI) computations (Sections~\ref{subsec:result_cliques}, \ref{subsec:result_patterns}, and \ref{subsec:result_iso}).
We also discuss the impact of the result set size ($k$) on subgraph discovery computations (Section~\ref{subsec:result_k}) and the performance of our virtual priority queue implementation (Section~\ref{subsec:result_queue}).

\subsection{Experimental setup} 
\label{subsec:setup}

\begin{table}[t]
\centering
\caption{Datasets}
\label{table:datasets}
\begin{tabular}{c c c c} \hline
&$|V|$ & $|E|$ & distinct labels\\ \hline
\texttt{Email}~\cite{leskovec2007graph} & 986 &16k &-\\
\texttt{CiteSeer}~\cite{yan2004graph} & 3.3k & 4.5k &6\\
\texttt{MiCo}~\cite{yan2004graph} & 100k & 1.1m & 29 \\ 
\texttt{YouTube}~\cite{yang2015defining} & 1.1m  &2.9m &-\\ 
\texttt{Patents}~\cite{hall.url11.patent} & 2.7m &14m &37\\ 
\hline\end{tabular}
\end{table}

\noindent{\bf Datasets.}
We employ five graph datasets from diverse domains and at different scales as shown in Table~\ref{table:datasets}.
Vertices in the \texttt{Email} dataset represent people and each edge between two vertices indicates that at least one email message was sent between the people corresponding to the vertices.
The \texttt{CiteSeer} dataset represents a citation network in which each publication is labeled by its research area.  
The \texttt{MiCo} dataset expresses a co-authorship network where authors are labeled by their research interests and a pair of authors is connected if they co-authored at least one publication.
The \texttt{YouTube} and \texttt{Patents} datasets represent a social network among the users of the service and a citation network among the US patents for the period between $1963$ to $1999$. In the \texttt{Patents} dataset, each vertex is labeled by the year in which the patent was granted.

\noindent
{\bf Systems Compared.} 
We compare two versions of our system, \texttt{Nuri} (supporting targeted expansion, pruning, and prioritization) and \texttt{Nuri-NP} (supporting only targeted expansion), and Arabesque~\cite{teixeira.sosp15.arabesque} as a representative of the prior subgraph discovery systems~\cite{teixeira.sosp15.arabesque, quamar2014nscale, chen2018g}.
In contrast to NScale~\cite{quamar2014nscale} and GMiner~\cite{chen2018g}, Arabesque supports aggregate subgraph discovery (e.g., pattern mining) and its implementation is openly available. 

Both \texttt{Nuri} and \texttt{Nuri-NP} are implemented as single-threaded Java programs using the standard Java $8$ distribution.
In our experiments, each of these versions was run on a {\em single core} at 1.90GHz of an Intel(R) Xeon(R) E5-4640 server ($256GB$ memory) running Red Hat Enterprise Linux Server release $7.5$. 
In contrast, Arabesque was executed on a {\em total of $40$ cores} from $5$ Intel(R) Xeon(R) E5430 servers (each with $8$ cores at 2.66GHz and $16GB$ memory), benefiting from its distributed computing capability.
In this section, the results obtained from Arabesque are labeled ``\texttt{Abq40}''.

\noindent
{\bf Evaluation Metrics.} 
We quantify the performance of both \oursyst and Arabesque by the following metrics:
(1) {\em number of candidate subgraphs}: Both Arabesque and \oursyst examine candidate subgraphs obtained through expansion until the desired result is obtained.
Hence, we consider the number of candidate subgraphs as the basic cost unit, allowing us to represent the inherent computational load without being affected by implementation details (e.g., specific data structures used to represent subgraphs and the actual implementation code for grouping subgraphs according to their patterns).
(2) {\em completion time of subgraph discovery}: 
This metric allows us to compare different systems/techniques from the user's point of view.
It also compensates for the limitation of the first metric, which cannot incorporate the overhead for optimization (e.g., time spent for prioritizing subgraphs and evaluating pruning conditions). 
In addition, it allows us to take into account Arabesque's ability to speed up subgraph discovery through distributed computing.

\begin{figure*}[ht]
\footnotesize

\begin{minipage}[b]{0.33\linewidth}
     \subfloat[subgraphs\label{fig:result_pattern_mining_patent_subgraphs}]{%
       \includegraphics[width=0.48\textwidth, trim={1.6cm 0.4 .6cm 0}]{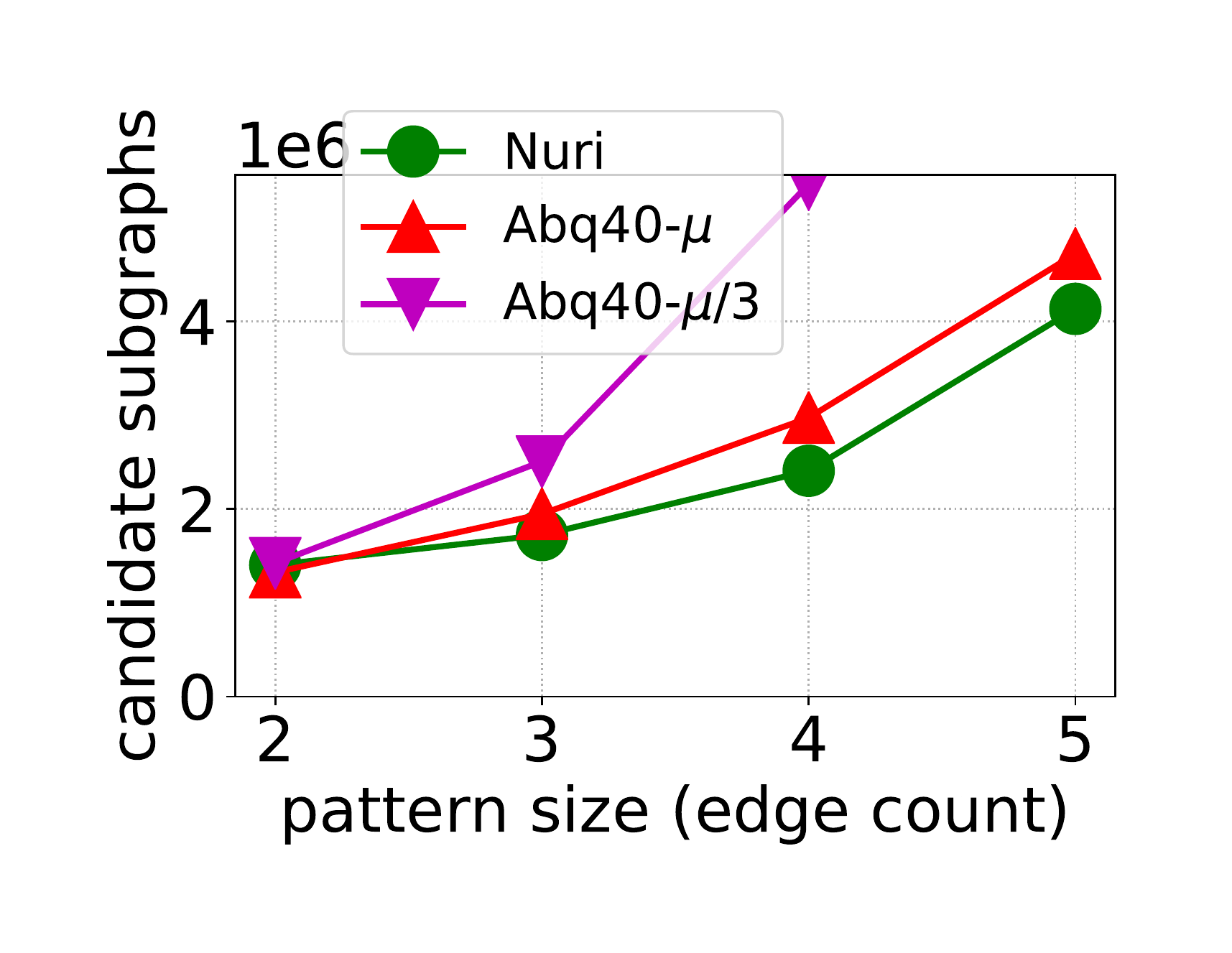}
     }
     \subfloat[time\label{fig:result_pattern_mining_patent_time}
     ]{%
       \includegraphics[width=0.48\textwidth, trim={1.6cm 0 .6cm 0}]{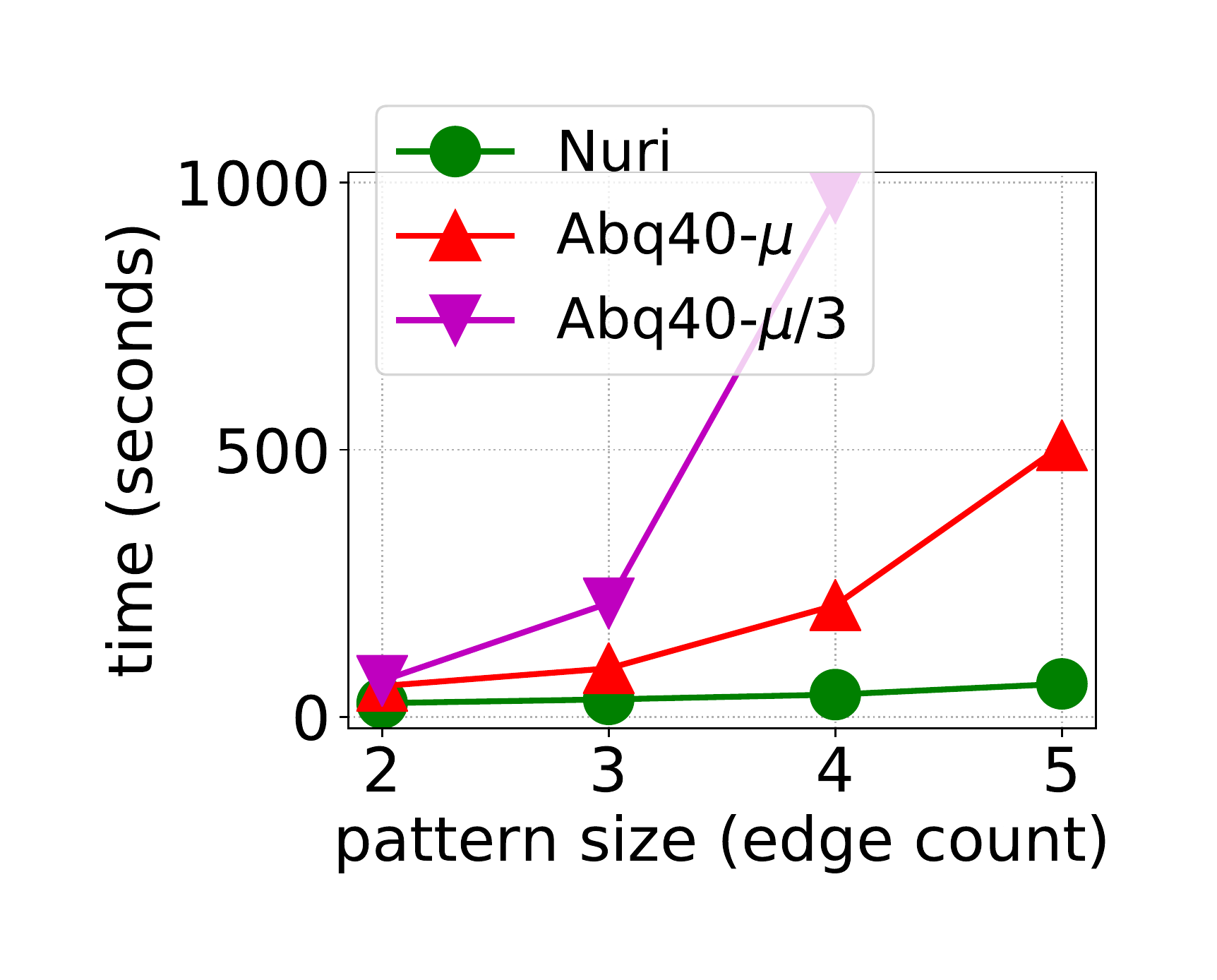}
     }
    \caption{Pattern Mining (\texttt{Patents})}\label{fig:result_pattern_mining_patent}
\end{minipage}
\begin{minipage}[b]{0.33\linewidth}
     \subfloat[subgraphs\label{fig:result_pattern_mining_CiteSeer_subgraphs}]{%
       \includegraphics[width=0.48\textwidth, trim={1.6cm 0 .6cm 0}]{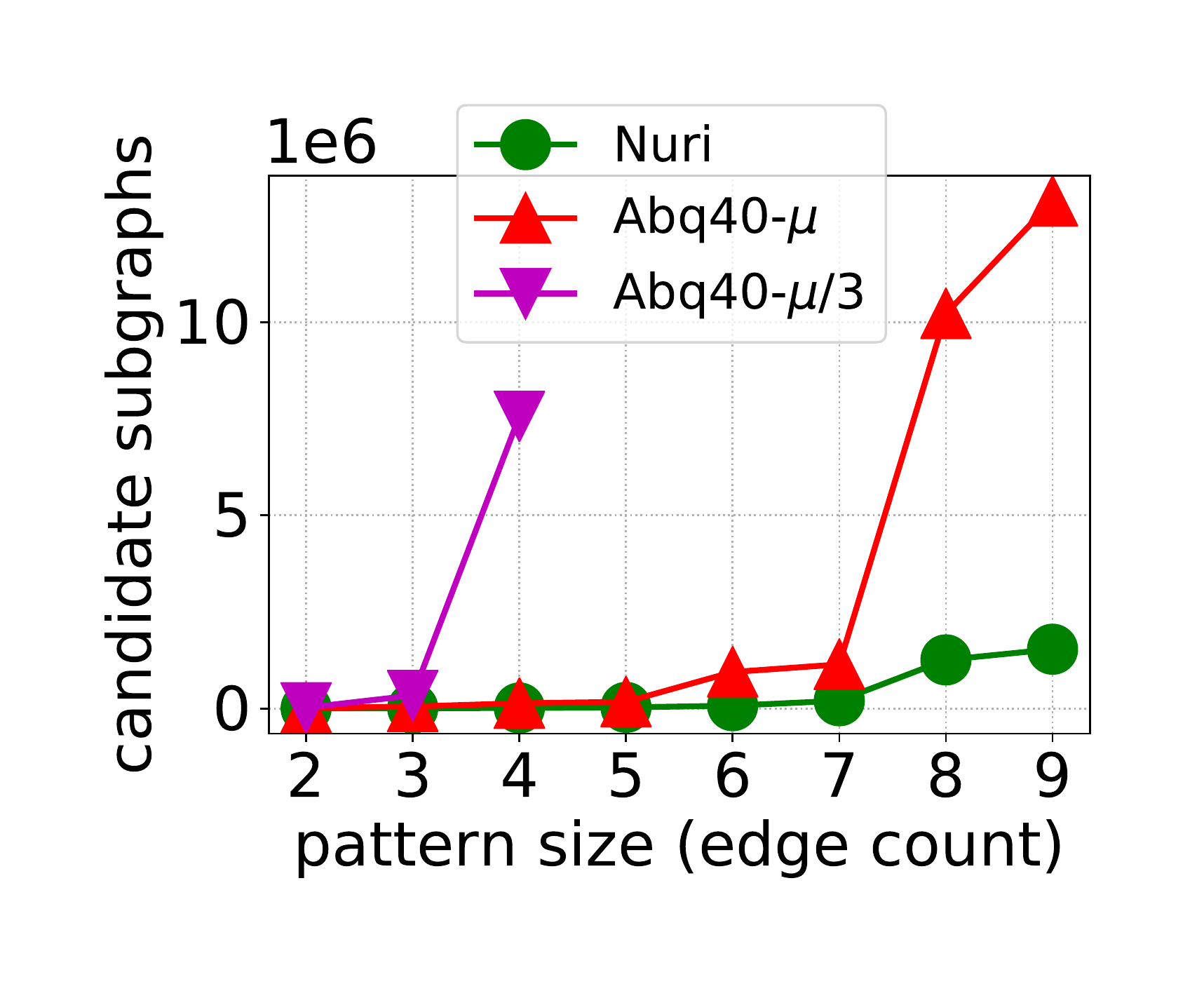}
     }
     \subfloat[time\label{fig:result_pattern_mining_CiteSeer_time}
     ]{%
       \includegraphics[width=0.48\textwidth, trim={1.6cm 0 .6cm 0}]{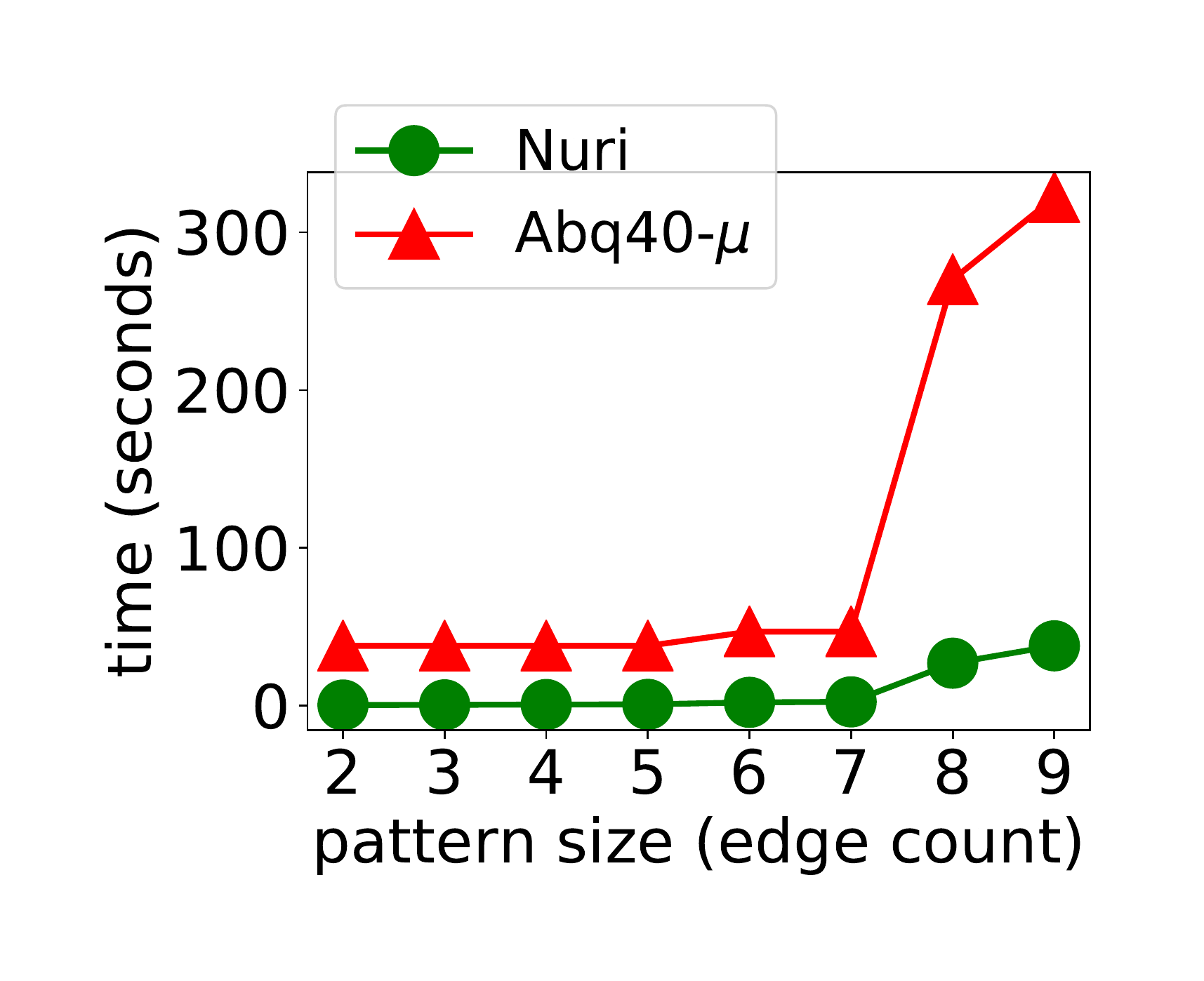}
     }
    \caption{Pattern Mining (\texttt{CiteSeer})}\label{fig:result_pattern_mining_CiteSeer}
\end{minipage}
\begin{minipage}[b]{0.33\linewidth}
     \subfloat[subgraphs\label{fig:result_pattern_mining_MiCo_subgraphs}]{%
       \includegraphics[width=0.48\textwidth, trim={1.6cm 0 .6cm 0}]{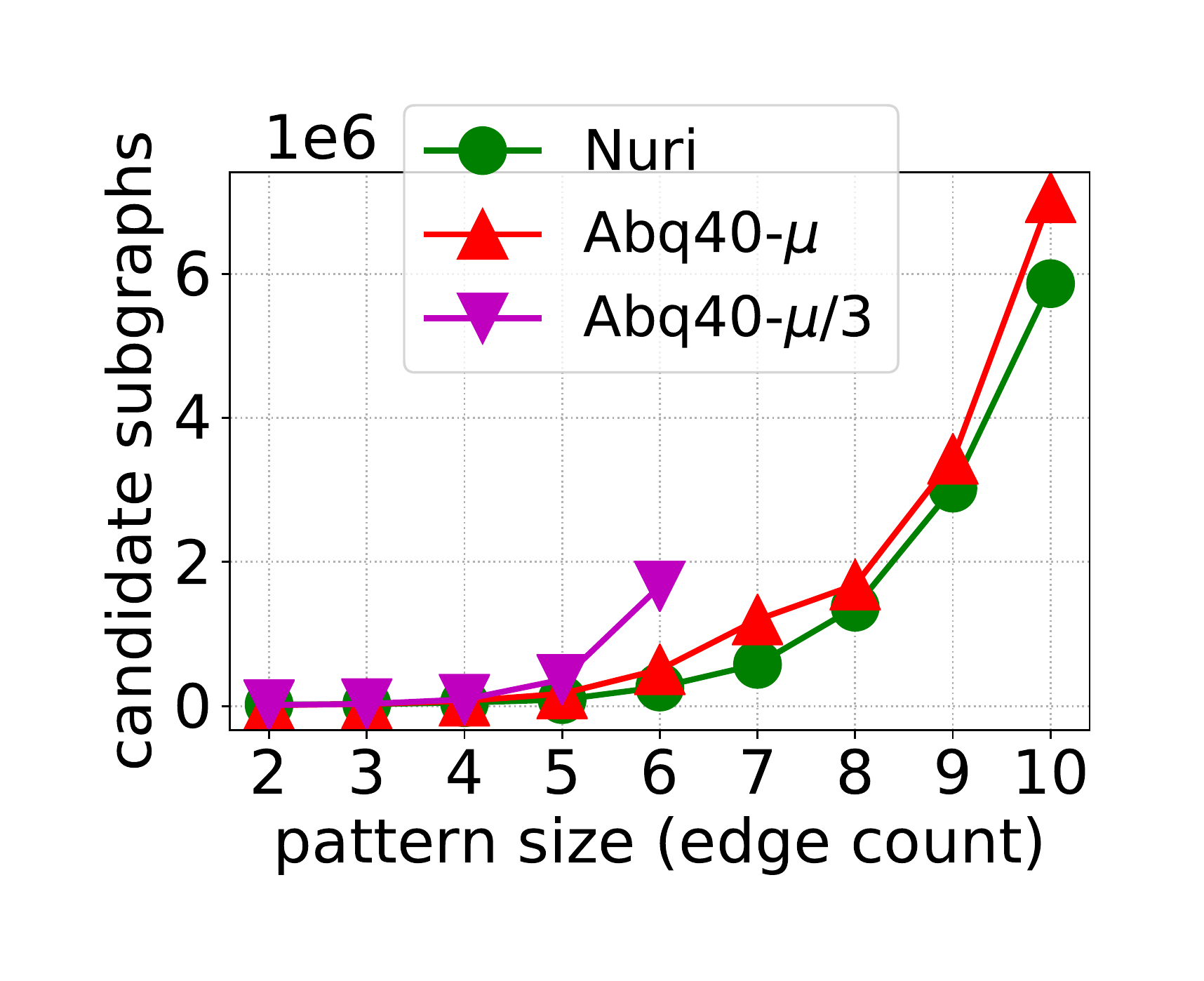}
     }
     \subfloat[time\label{fig:result_pattern_mining_MiCo_time}
     ]{%
       \includegraphics[width=0.48\textwidth, trim={1.6cm 0 .6cm 0}]{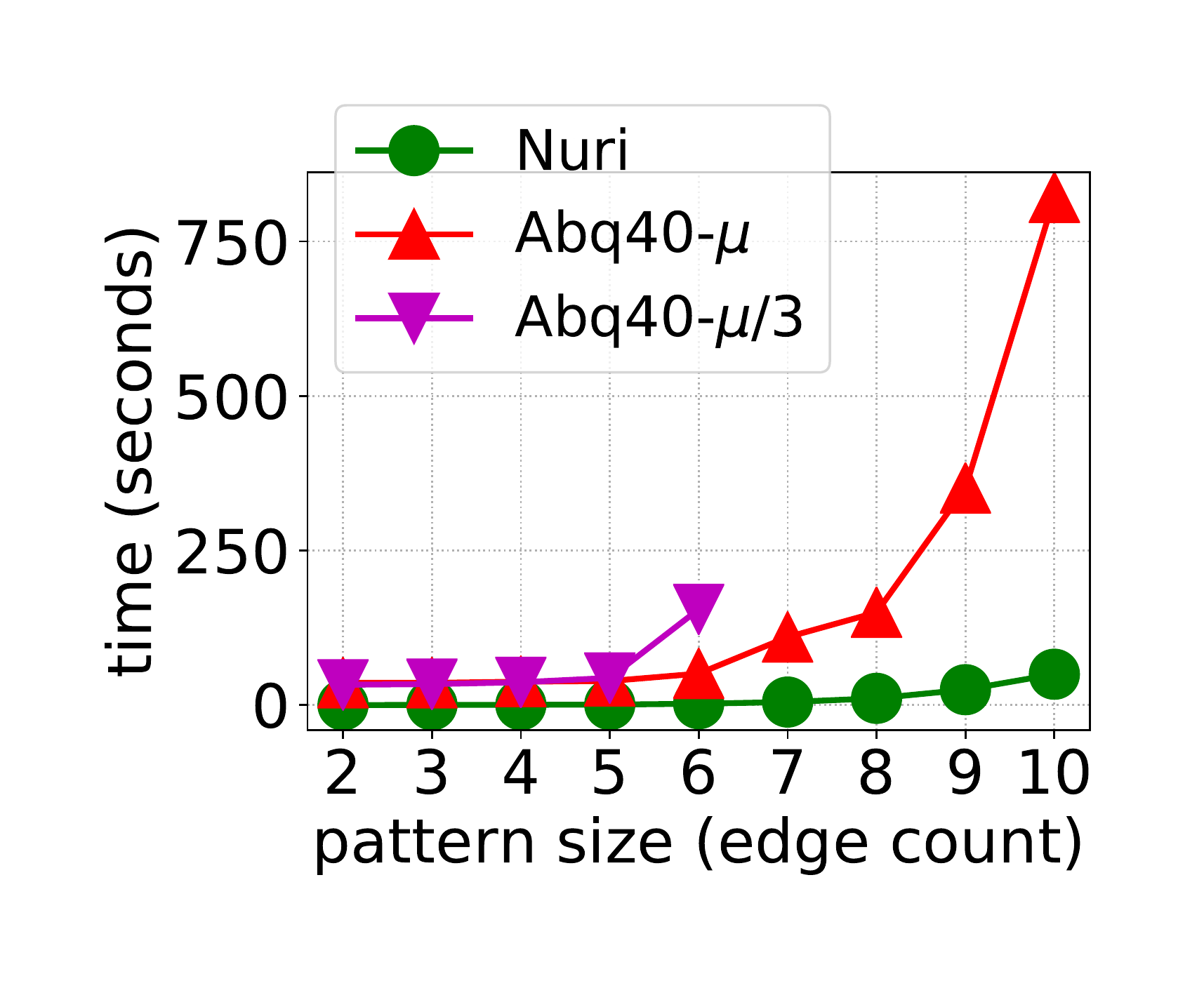}
     }
    \caption{Pattern Mining (\texttt{MiCo})}\label{fig:result_pattern_mining_MiCo}
\end{minipage}

\end{figure*}

\subsection{Clique Discovery Evaluation}
\label{subsec:result_cliques}

To obtain clique discovery (CD) results from Arabesque, we instrumented it to find all cliques and then select the largest clique(s) among them\footnote{The original implmentation supports only clique enumeration at a predefined size.}. 
We created increasingly denser data graphs using the \texttt{Email}, \texttt{MiCo}, and \texttt{Youtube} datasets by repeatedly adding batches of randomly chosen edges to an empty graph.
Denser graphs tend to include more and larger cliques, increasing the complexity of clique discovery.

Figures~\ref{fig:result_cliques_email}, ~\ref{fig:result_cliques_mico}, and ~\ref{fig:result_cliques_youtube} show the CD results obtained for the three datasets. As expected, both the number of candidate subgraphs and completion time increase with the density of the data graph for all of the systems. 
In Figure~\ref{fig:result_cliques_email}, the gap between \texttt{Nuri-NP} and \texttt{Abq40} is due to the advantage of {\em targeted expansion}, which allows \oursyst to explore only relevant subgraphs (cliques).
On the other hand, Arabesque exhaustively creates subgraphs and then filters out irrelevant (non-clique) subgraphs.
The difference between \texttt{Nuri-NP} and \texttt{Nuri} demonstrates that \oursyst can safely ignore a large number of candidate subgraphs ({\em pruning}) benefiting from {\em prioritization} (for $16k$ edges, \texttt{Nuri} examines only $1/26$ of subgraphs and is $29$x faster compared to \texttt{Nuri-NP}).
It is also evident that the benefits of pruning and prioritization increase with density.
In contrast to \oursyst, Arabesque must find all smaller subgraphs before any larger subgraph (no pruning/prioritization).
As a result, for data graphs with $15k$ edges, \oursyst can find the largest clique(s) by examining only a small fraction (about $1/400$) of subgraphs compared to Arabesque, resulting in $2$ orders of magnitude improvement in running time although \oursyst is given much less computing resources ($1$ core vs. $40$ cores).
For $16k$ edges, \oursyst completes the computation within $2$ minutes while Arabesque cannot within $10$ hours.

Figures~\ref{fig:result_cliques_mico} and \ref{fig:result_cliques_youtube} show the CD results obtained for the \texttt{MiCo} and \texttt{YouTube} datasets.
Given $500k$ edges from the \texttt{MiCo} dataset, \oursyst examines only $1/85$ of subgraphs compared to Arabesque (Figure~\ref{fig:result_cliques_MiCo_candidates}), resulting in a $65$x improvement in running time (Figure ~\ref{fig:result_cliques_MiCo_time}). 
On $600k$ edges from the \texttt{MiCo} dataset, neither Arabesque nor \oursyst (without pruning and prioritization) finish their computation within $10$ hours whereas \oursyst (with pruning and prioritization) completes within $47$ minutes.
Given $2.8M$ edges from the \texttt{YouTube} dataset, compared to Arabesque, \oursyst examines $2$ orders of magnitude fewer candidate subgraphs (Figure ~\ref{fig:result_cliques_youtube_candidates}) and is 1 order of magnitude faster (Figure~\ref{fig:result_cliques_youtube_time}).
When $2.9M$ edges are used, Arabesque does not complete within $10$ hours while \oursyst finishes in less than $1$ hour.

\begin{figure*}[ht]
\footnotesize

\begin{minipage}[b]{0.33\linewidth}
     \subfloat[subgraphs\label{fig:result_iso_CiteSeer_candidates}]{%
       \includegraphics[width=0.48\textwidth, trim={1.6cm 0.4 0 0}]{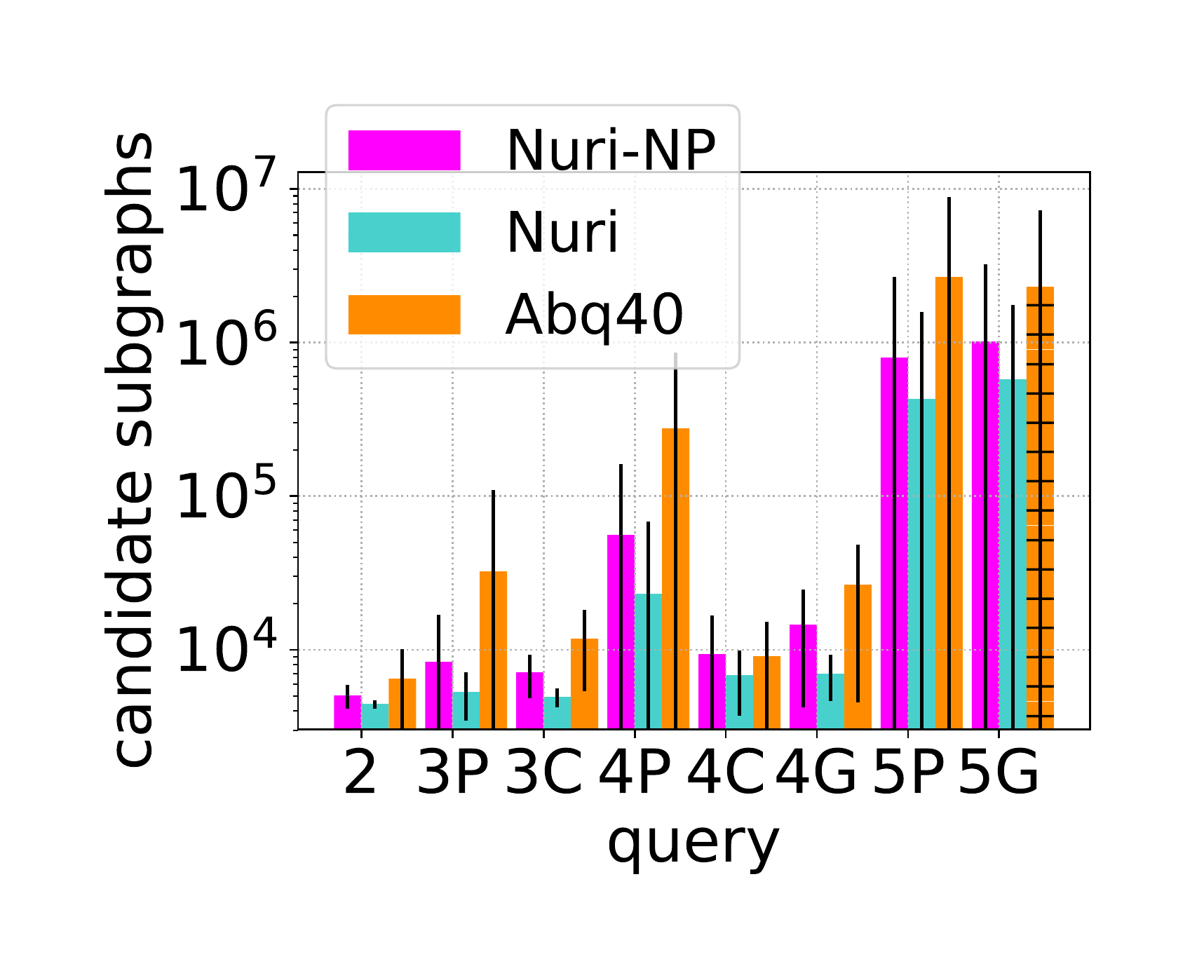}
     }
     \subfloat[time\label{fig:result_iso_CiteSeer_time}
     ]{%
       \includegraphics[width=0.48\textwidth, trim={1.6cm 0 0 0}]{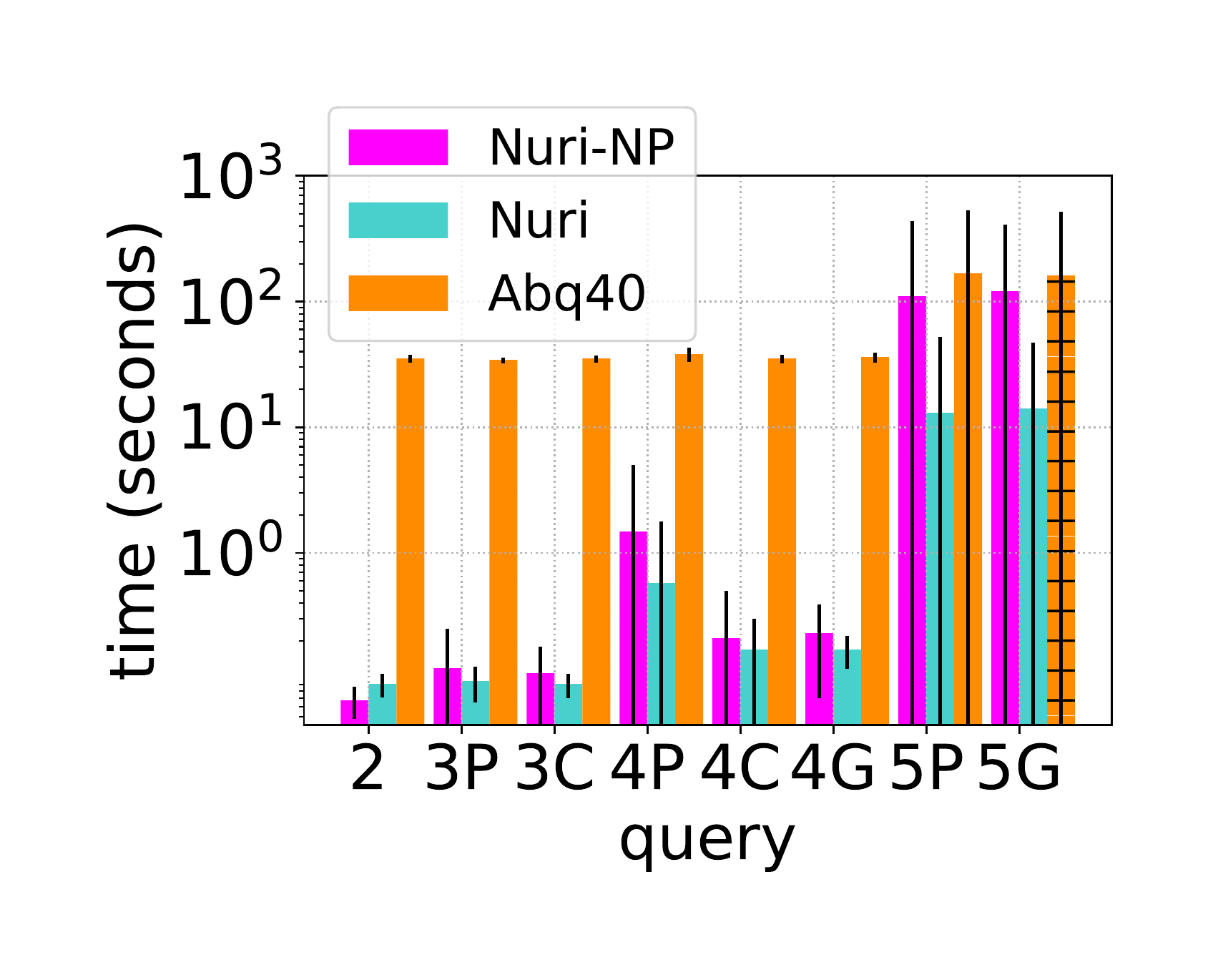}
     }
    \caption{SI (\texttt{CiteSeer})}\label{fig:result_iso_CiteSeer}
\end{minipage}
\begin{minipage}[b]{0.33\linewidth}
     \subfloat[subgraphs\label{fig:result_iso_MiCo_candidates}]{%
       \includegraphics[width=0.48\textwidth, trim={1.6cm 0 0 0}]{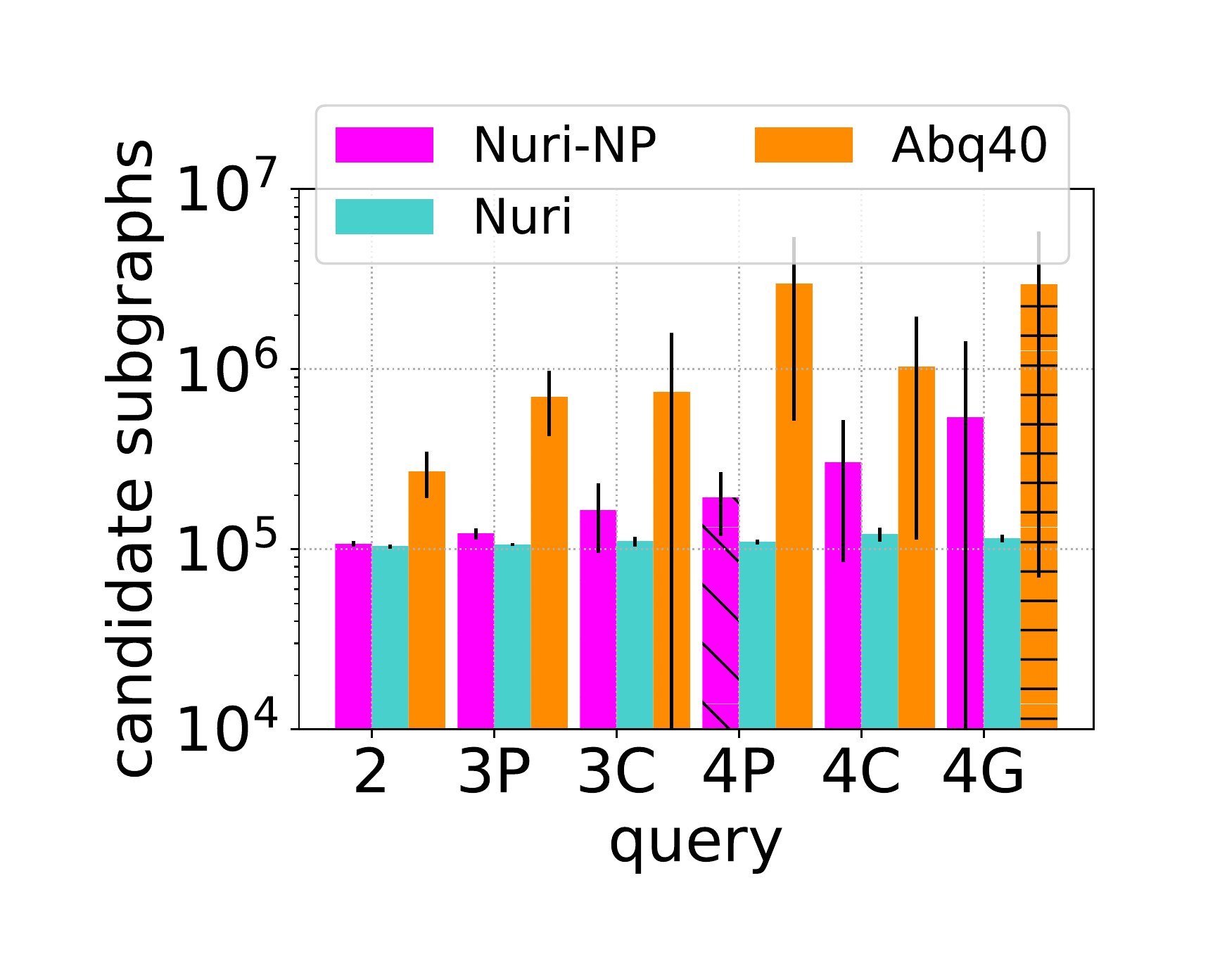}
     }
     \subfloat[time\label{fig:result_iso_MiCo_time}
     ]{%
       \includegraphics[width=0.48\textwidth, trim={1.6cm 0 0 0}]{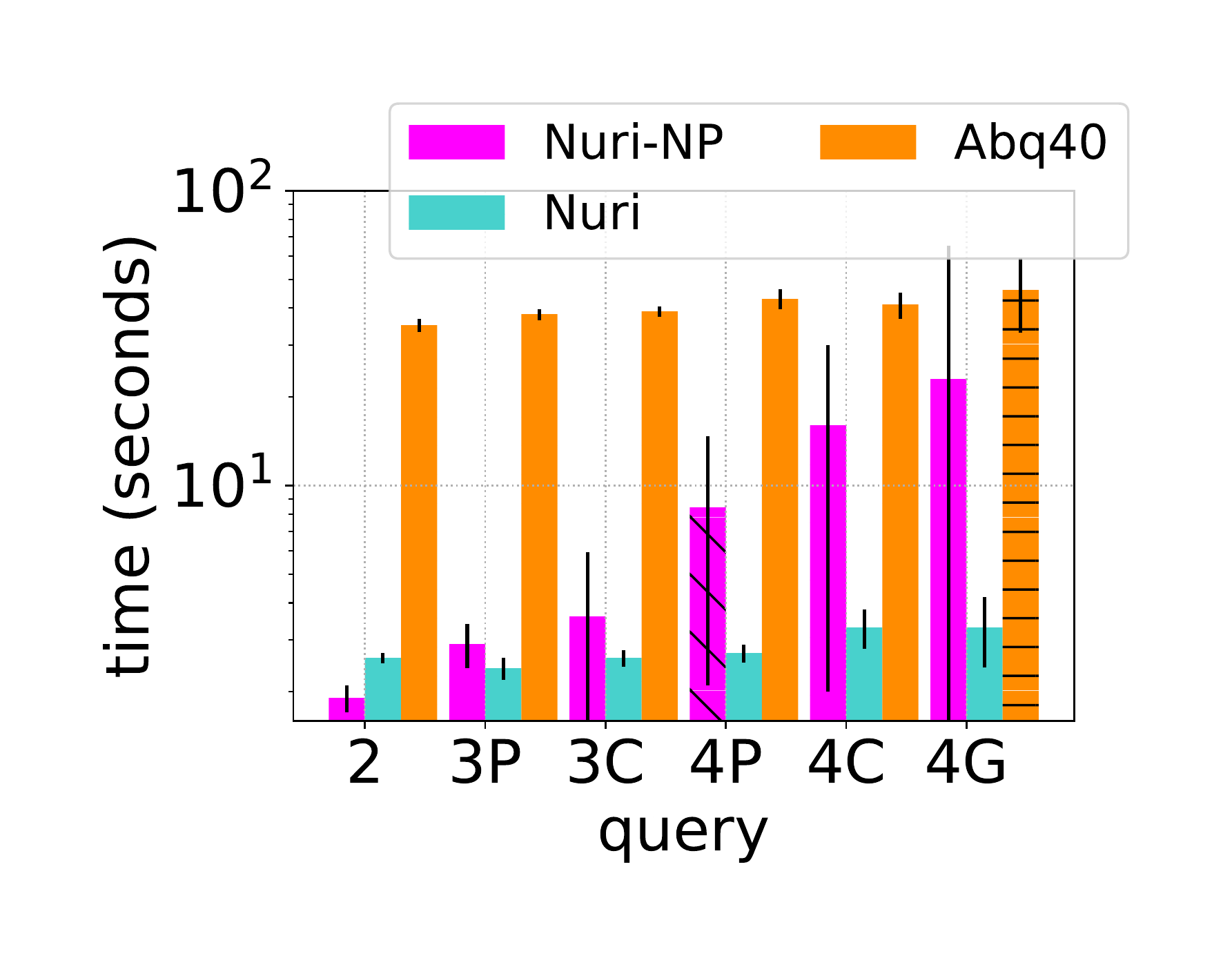}
     }
    \caption{SI (\texttt{MiCo})}\label{fig:result_iso_MiCo}
\end{minipage}
\begin{minipage}[b]{0.33\linewidth}
     \subfloat[subgraphs\label{fig:isomorphism_effect_selectivity_candidates}]{%
       \includegraphics[width=0.48\textwidth, trim={1cm 0 .6cm 0}]{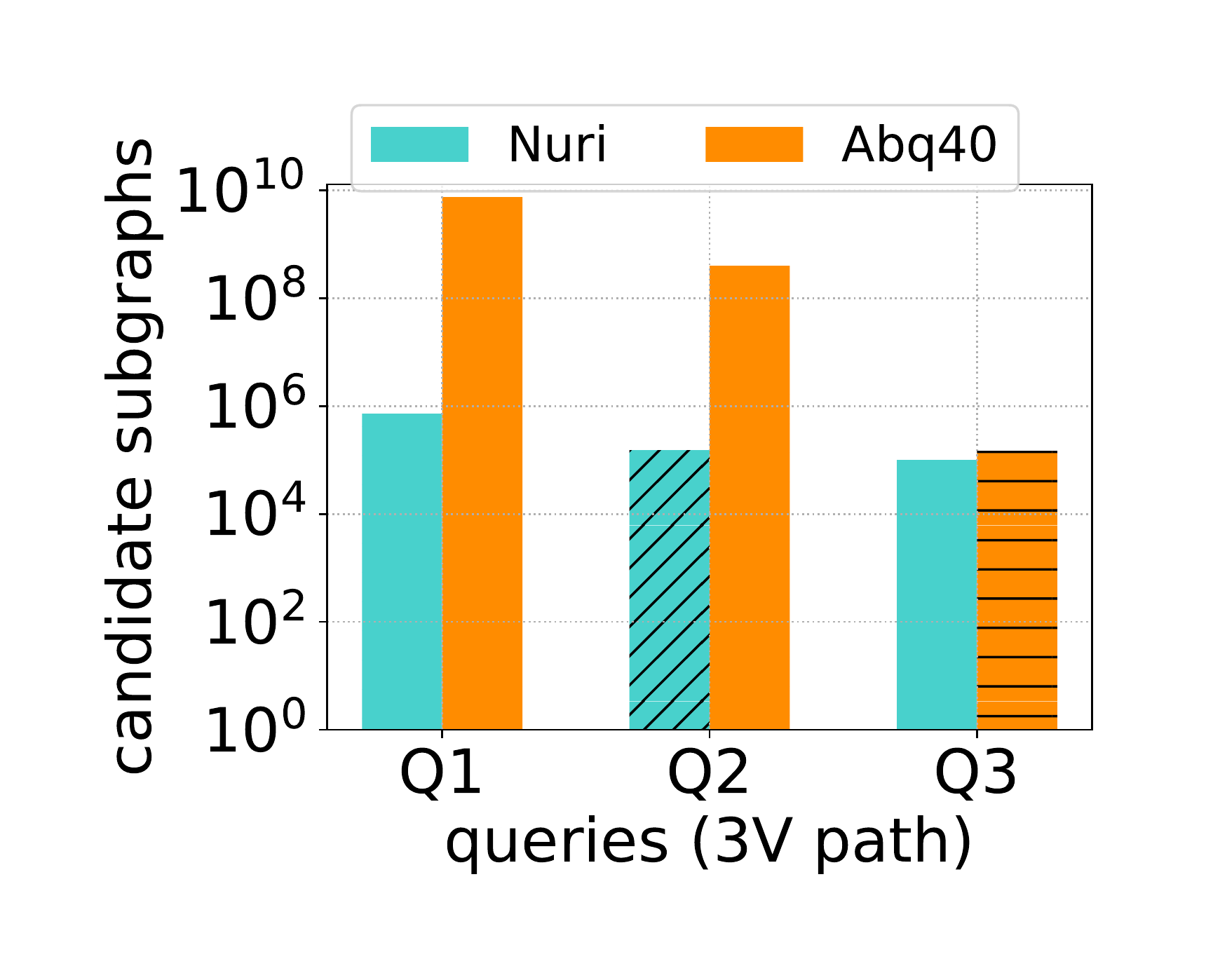}
     }
     \hfill
     \subfloat[time\label{fig:isomorphism_effect_selectivity_time}
     ]{%
       \includegraphics[width=0.48\textwidth, trim={1cm 0 .6cm 0}]{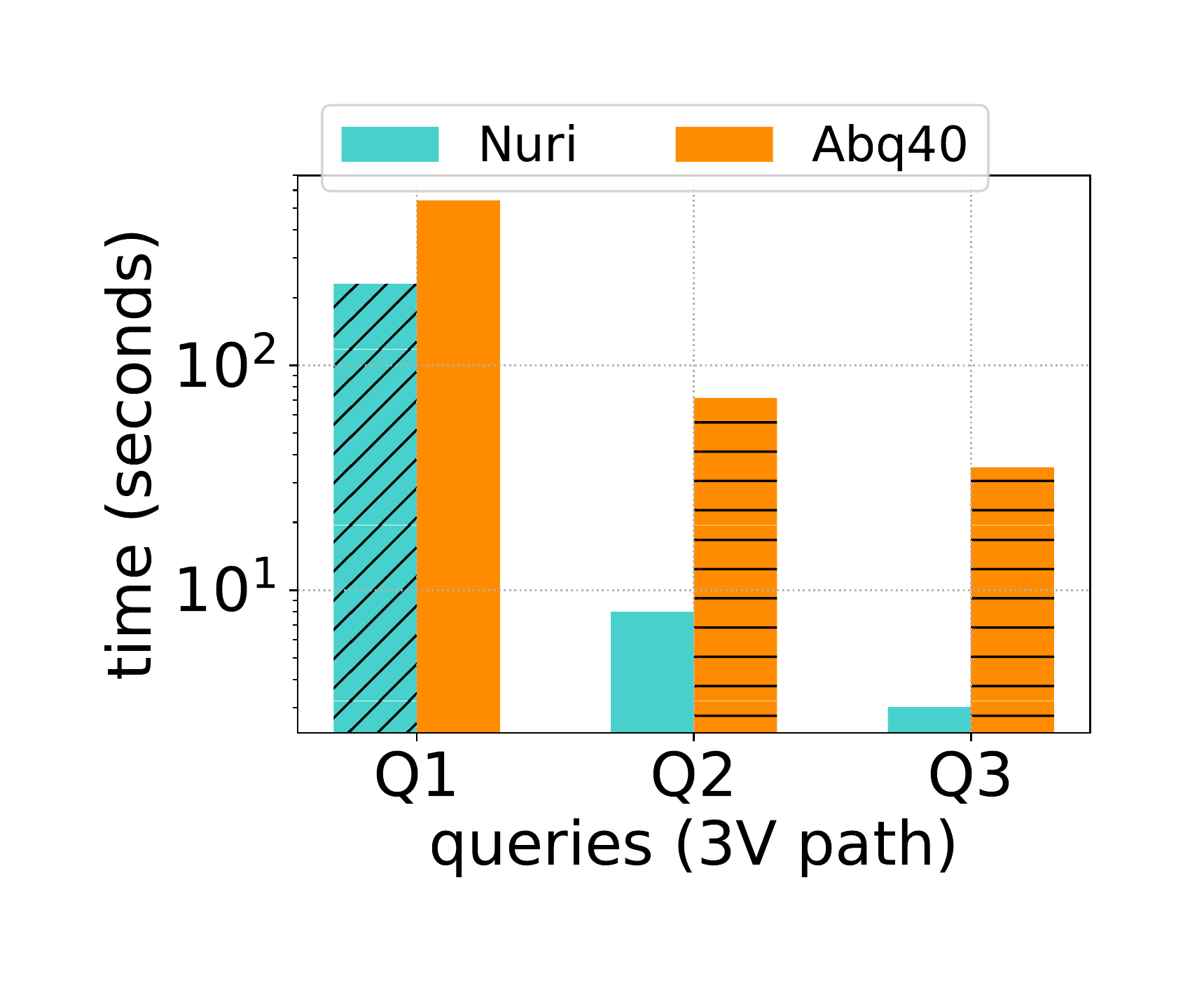}
     }
    \caption{Impact of Selectivity
    }\label{fig:isomorphism_effect_selectivity}
\end{minipage}

\end{figure*}

\subsection{Pattern Mining Evaluation}
\label{subsec:result_patterns}

Our pattern mining (PM) implementation finds, given a number $M$, the most frequent patterns of size $M$ (i.e., $M$-edge patterns) in the data graph (Listing~\ref{listing:patterns}).
To obtain the same result, we instrumented Arabesque to (1) find all of the $M$-edge patterns whose frequency is no lower than a threshold $T$ and then (2) select the most frequent one(s) among these patterns.
This Arabesque implementation has the benefit of proactively pruning out (i.e., not expanding) subgraphs whose pattern has a frequency lower than $T$ (due to anti-monotonicity~\cite{bringmann.pakdd08.minimImageBased}, expansions of these subgraphs cannot lead to the $M$-edge patterns whose frequency is at least $T$).
In real-world use cases, however, it is difficult to appropriately set $T$ for Arabesque since the maximum frequency (denoted $\mu$) over patterns of size $M$ is {\em not known in advance}.
If $T$ is set to a value greater than $\mu$, Arabesque cannot report any patterns of size $M$ since all of the patterns of size $M$ have a frequency lower than $T$.
If $T$ is assigned a value lower than $\mu$, it examines subgraphs that are unnecessary for the purpose of finding the most frequent pattern(s).

Figure~\ref{fig:result_pattern_mining_patent} shows the PM result obtained for the \texttt{Patents} dataset.
When the threshold $T$ for Arabesque is set to $\mu$, \oursyst and Arabesque explore a similar number of subgraphs (\texttt{\oursyst} and \texttt{Abq40-$\mu$} in Figure \ref{fig:result_pattern_mining_patent_subgraphs}).
The difference in the number of subgraphs between these systems is due to their use of different expansion approaches (Section~\ref{subsec:aggregate}).
In Figure \ref{fig:result_pattern_mining_patent_time}, the gap between \texttt{\oursyst} and \texttt{Abq40-$\mu$} shows the benefits of pattern-oriented expansion (\oursyst) over edge-oriented expansion (Arabesque).
Under pattern-oriented expansion (Section~\ref{subsec:aggregate}), when subgraph $s$ expands into $s'$, the pattern of $s'$ can be quickly obtained by appending only one edge to the pattern of $s$ ({\em incremental pattern generation}). 
On the other hand, under edge-oriented expansion, the pattern of each subgraph is always computed from scratch by converting that subgraph into its {\em canonical form} with high overhead~\cite{teixeira.sosp15.arabesque}.
When the threshold $T$ for Arabesque is set to $\mu/3$, Arabesque examines subgraphs unnecessary for the purpose of finding the most frequent patterns, in contrast to \oursyst which automatically prunes out such subgraphs.
For $M=4$ and $T=\mu/3$, Arabesque explores 2.5x more subgraphs than \oursyst (Figure \ref{fig:result_pattern_mining_patent_subgraphs}).
For $M=5$ and $T=\mu/3$, Arabesque does not complete its operation as its memory requirement surpasses the capacity of the system.

Similar trends for PM can be seen for the \texttt{CiteSeer} and \texttt{MiCo} datasets (Figures~\ref{fig:result_pattern_mining_CiteSeer} and \ref{fig:result_pattern_mining_MiCo}, respectively).
In these datasets, only a few distinct labels are assigned to vertices, which allows an enormous number of subgraphs to match the same pattern (i.e., very high computational overhead).
For this reason, we introduced synthetic labels, increasing the number of labels from $6$ to $48$ in \texttt{CiteSeer}  and from $29$ to $87$ in \texttt{MiCo} (i.e., reduced computational overhead).
In Figures~\ref{fig:result_pattern_mining_CiteSeer} and \ref{fig:result_pattern_mining_MiCo}, Arabesque explores much more subgraphs compared to \oursyst, incurring substantially higher overhead.
The benefits of \oursyst over Arabesque ({\em pruning}  and {\em prioritization}) become more evident as the pattern size increases.
When $T$ is set to $\mu/3$, Arabesque does not complete due to its high memory demand when the pattern size is 4 for \texttt{CiteSeer} and 6 for \texttt{MiCo}.

\subsection{Subgraph Isomorphism Evaluation}
\label{subsec:result_iso}

Our top-$k$ subgraph isomorphism (SI) implementation discovers, in the data graph, the $k$ highest-scored subgraphs that are isomorphic to a given query graph, where the score of each subgraph is defined as the sum of the degree of the vertices in that subgraph (Section~\ref{subsec:api_indexing}).
This implementation adopts Ullman's algorithm~\cite{ullmann.j1976.iso} to efficiently find the subgraphs that match the query graph ({\em targeted expansion}).
Since Arabesque~\cite{teixeira.sosp15.arabesque} does not support targeted expansion, our SI implementation for Arabesque exhaustively explores subgraphs while filtering out subgraphs that do not pass a subgraph isomorphism test (a user-provided implementation).
We also extended Arabesque so that it can maintain the $k$ highest-scored subgraphs.

To conduct SI computations, we pre-computed query graphs of sizes from 2 to 5 by running a sampling algorithm \cite{li.icde15.rcmh} on each data graph constructed from the \texttt{CiteSeer} and \texttt{MiCo} datasets.
For 4-vertex subgraphs, we considered three different types (path, clique, and general that are labeled ``\texttt{4P}'', ``\texttt{4C}'', and ``\texttt{4G}'').
We also considered these three types for 5-vertex subgraphs.
Since $3$-vertex subgraphs can only form a clique or a path, we did not include the general subgraph type.
For 2-vertex subgraphs, we considered only one type (labeled ``\texttt{2}'') which corresponds to both the clique and path types.
For each type of query graph, we carried out SI computations using 10 query graphs (i.e., samples) and then calculated the mean values for the number of subgraphs and completion time.
The graph from \texttt{CiteSeer} is small and sparse and thus contains very few cliques of size $5$.
Therefore, we did not consider cliques of size 5.

To facilitate pruning and prioritization, our SI implementation uses an index computed for every vertex in the data graph up to $d$-hops, where $d$ is the maximum diameter of all query graphs.
The index construction time for $4$-hops (sufficient for query graphs of up to size 5) was $1$ second for \texttt{CiteSeer} dataset. 
For the \texttt{MiCo} dataset, due to the size and the high density of the dataset, the index construction time for $3$-hops took $300$ minutes using a single core. 
Index construction (which is needed only once for multiple SI computations), however, is highly parallelizable because the computation for each vertex can be done independently. The index construction time was reduced to $600$ seconds when $32$ cores of a server were used.

As evident in Figures~\ref{fig:result_iso_CiteSeer_candidates} and \ref{fig:result_iso_MiCo_candidates}, as the query graph size increases, both Arabesque and \oursyst explore more subgraphs for each query type. 
In the \texttt{CiteSeer} graph, due to the sparsity of the graph, clique queries are very selective and in general explore a smaller number of subgraphs than path/general subgraph queries of the same size. 
The selectivity of queries also depends on the frequency of the query labels in the data graph, which affects the number of candidate subgraphs generated in the systems.
In Figures~\ref{fig:result_iso_CiteSeer_candidates} and \ref{fig:result_iso_MiCo_candidates}, the difference between \texttt{Nuri-NP} and \texttt{Abq40} demonstrates the benefits of targeted expansion.
Figures~\ref{fig:result_iso_CiteSeer_time} and \ref{fig:result_iso_MiCo_time} show that \oursyst is substantially faster than Arabesque taking advantage of pruning, prioritization, and targeted expansion even when it runs on a single core and Arabesque runs on a total of $40$ cores.

Figure~\ref{fig:isomorphism_effect_selectivity} illustrates how the selectivity of query graphs affects the SI computation for the \texttt{CiteSeer} dataset. 
In the figure, \texttt{Q1} is a non-selective query for which several million matches exist in the data graph. 
\texttt{Q2} is a mildly selective query and \texttt{Q3} is a highly selective query with fewer than $400$ matches in the data graph. 
Figure~\ref{fig:isomorphism_effect_selectivity_candidates} shows that \oursyst explores $4$ orders of magnitude fewer subgraphs than Arabesque for the non-selective query (\texttt{Q1}) due to prioritization and pruning and thus runs faster than Arabesque (Figure~\ref{fig:isomorphism_effect_selectivity_time}) despite using only 1/40 of computational resources compared to Arabesque.
Figure~\ref{fig:isomorphism_effect_selectivity_candidates} also shows that as the selectivity of query increases (\texttt{Q2} and \texttt{Q3}), the benefits of prioritization and pruning diminish since fewer subgraphs in the data graph match the query graph.
In Figure~\ref{fig:isomorphism_effect_selectivity_time}, Arabesque's time costs for \texttt{Q2} and \texttt{Q3} are mostly caused by its  distributed operation (particularly, coordination of multiple workers) rather than actual exploration of subgraphs.

\begin{figure*}[ht]
\footnotesize

\begin{minipage}[b]{0.66\linewidth}
     \subfloat[CD (\texttt{Email})\label{fig:effect_K_cliques}]{%
       \includegraphics[width=0.3\textwidth, trim={1.6cm 0.4 0 0}]{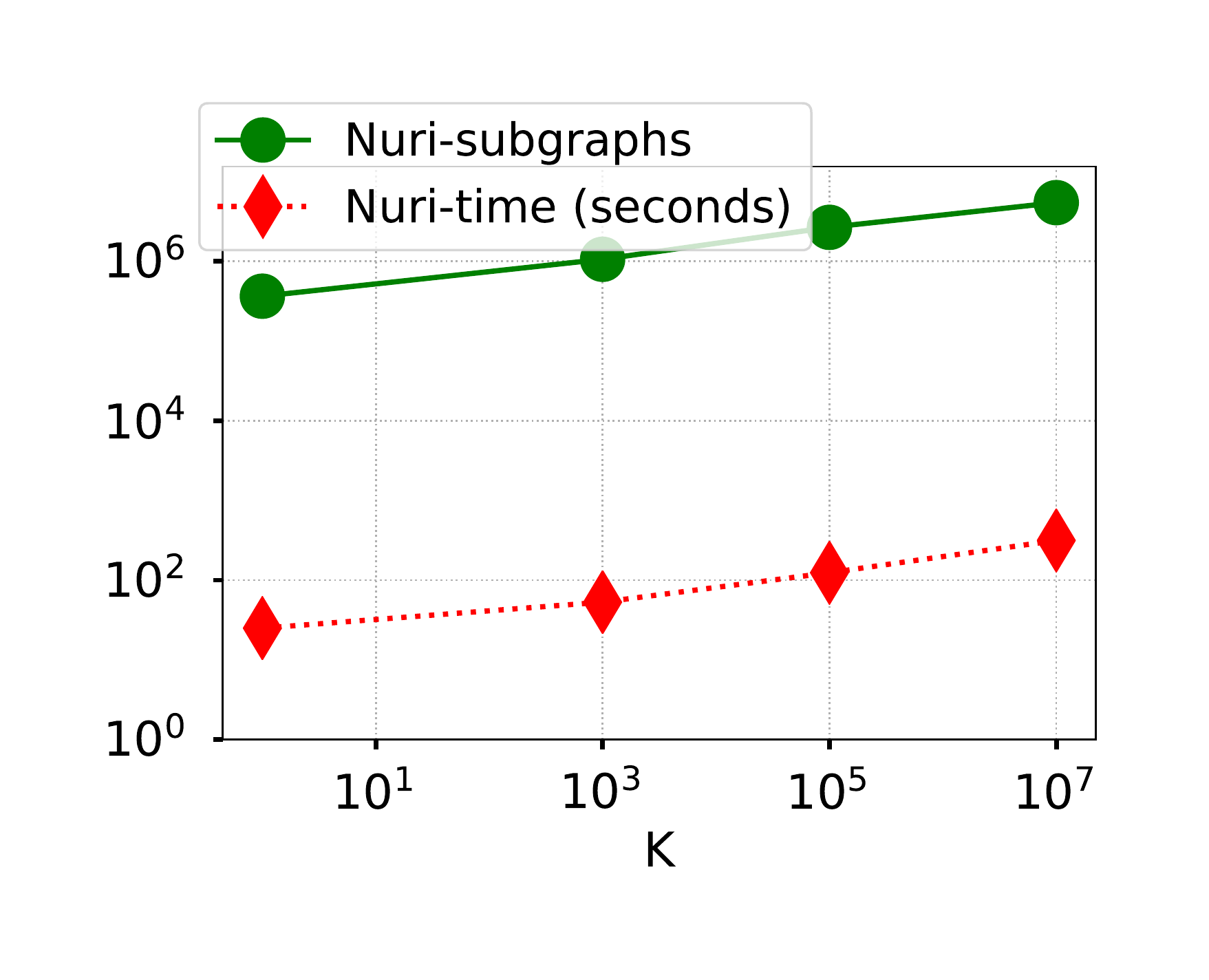}
     }
     \subfloat[PM (\texttt{CiteSeer})\label{fig:effect_K_PM_patents}
     ]{%
       \includegraphics[width=0.3\textwidth, trim={1.6cm 0 0 0}]{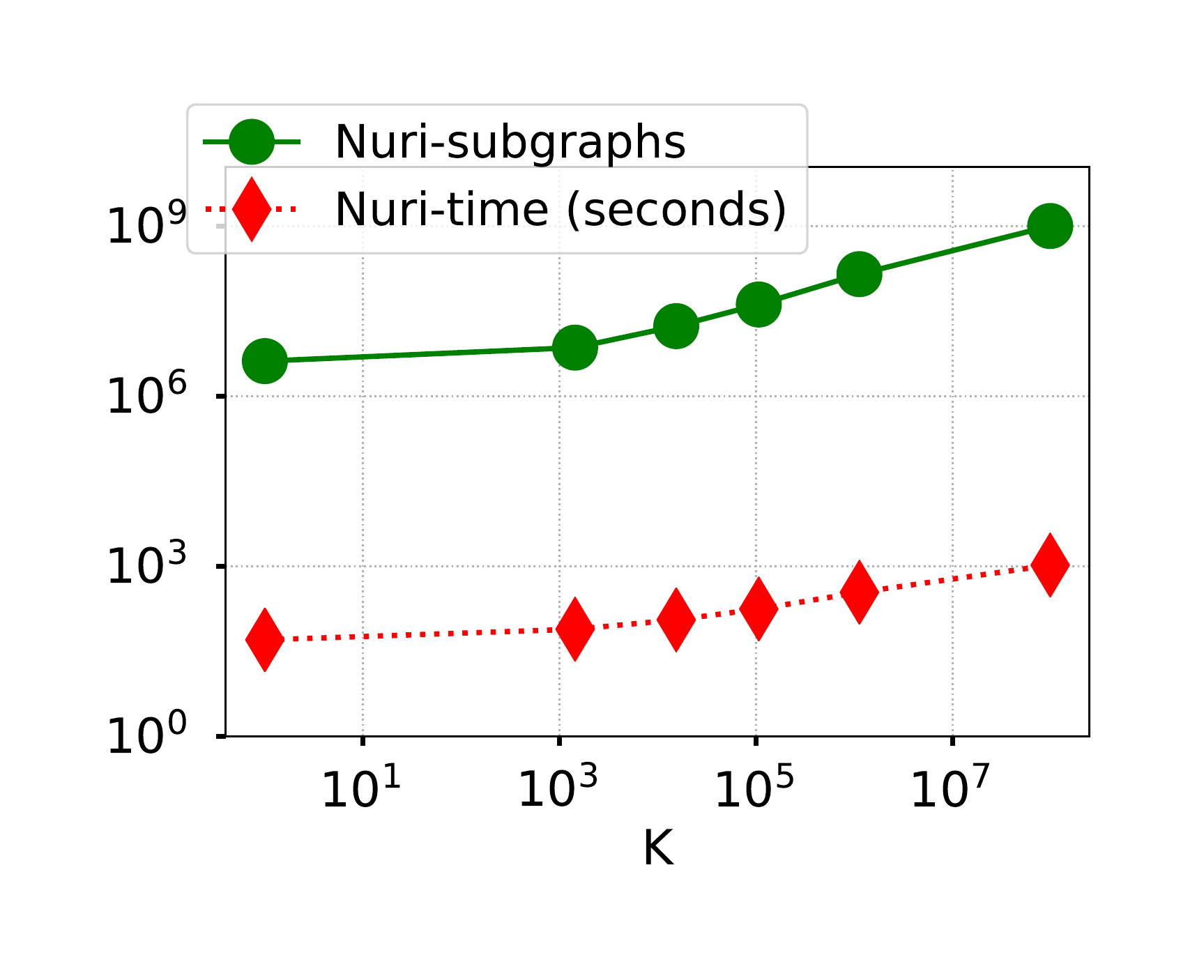}
     }
     \subfloat[SI (\texttt{CiteSeer})\label{fig:effect_K_ISO_CiteSeer}]{%
       \includegraphics[width=0.3\textwidth, trim={1.6cm 0 0 0}]{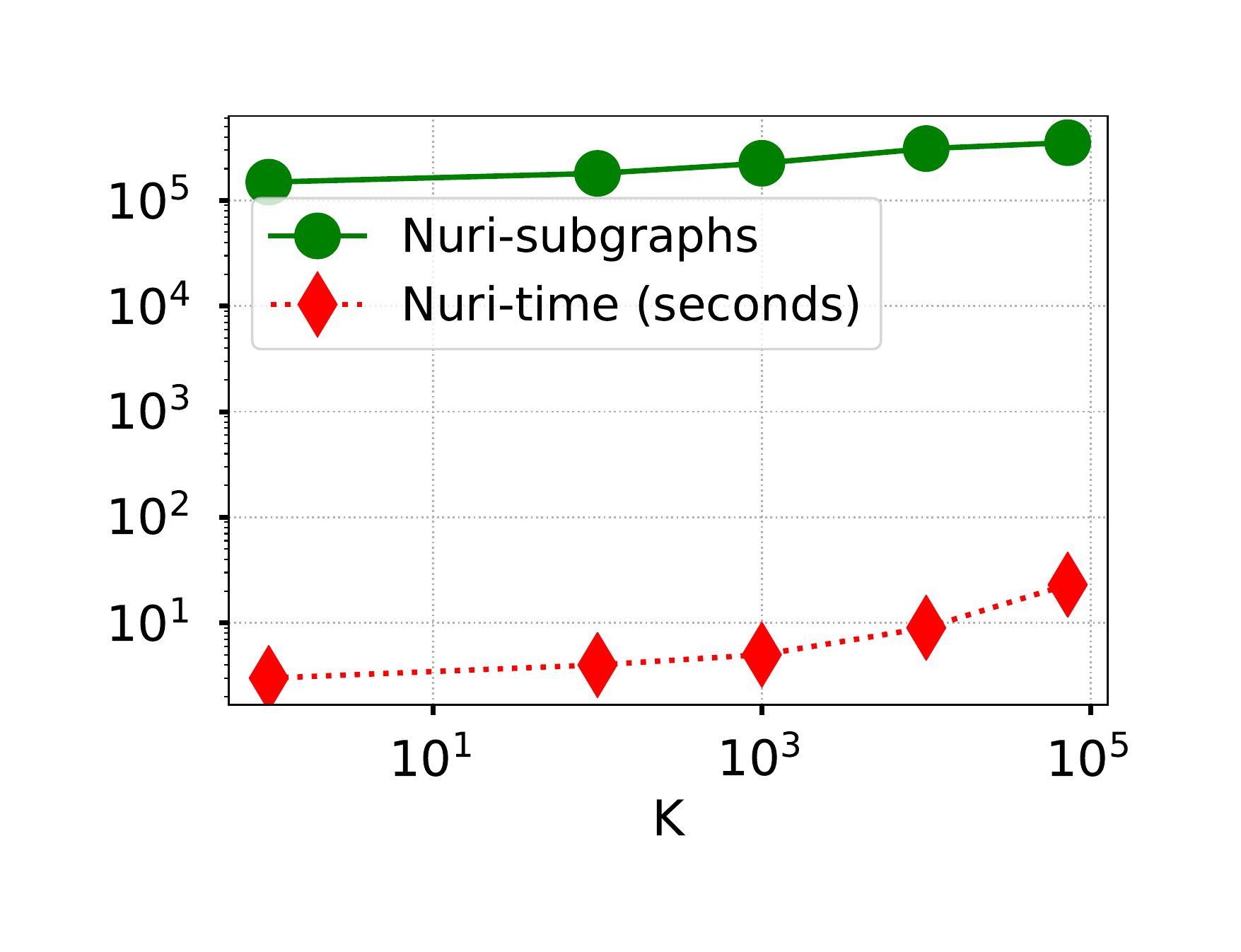}
     }
    \caption{Effect of Result Set Size ($k$)}\label{fig:topkeffect}
\end{minipage}
\begin{minipage}[b]{0.34\linewidth}
\includegraphics[width=.8\textwidth, trim={0 .5cm 0 0}]{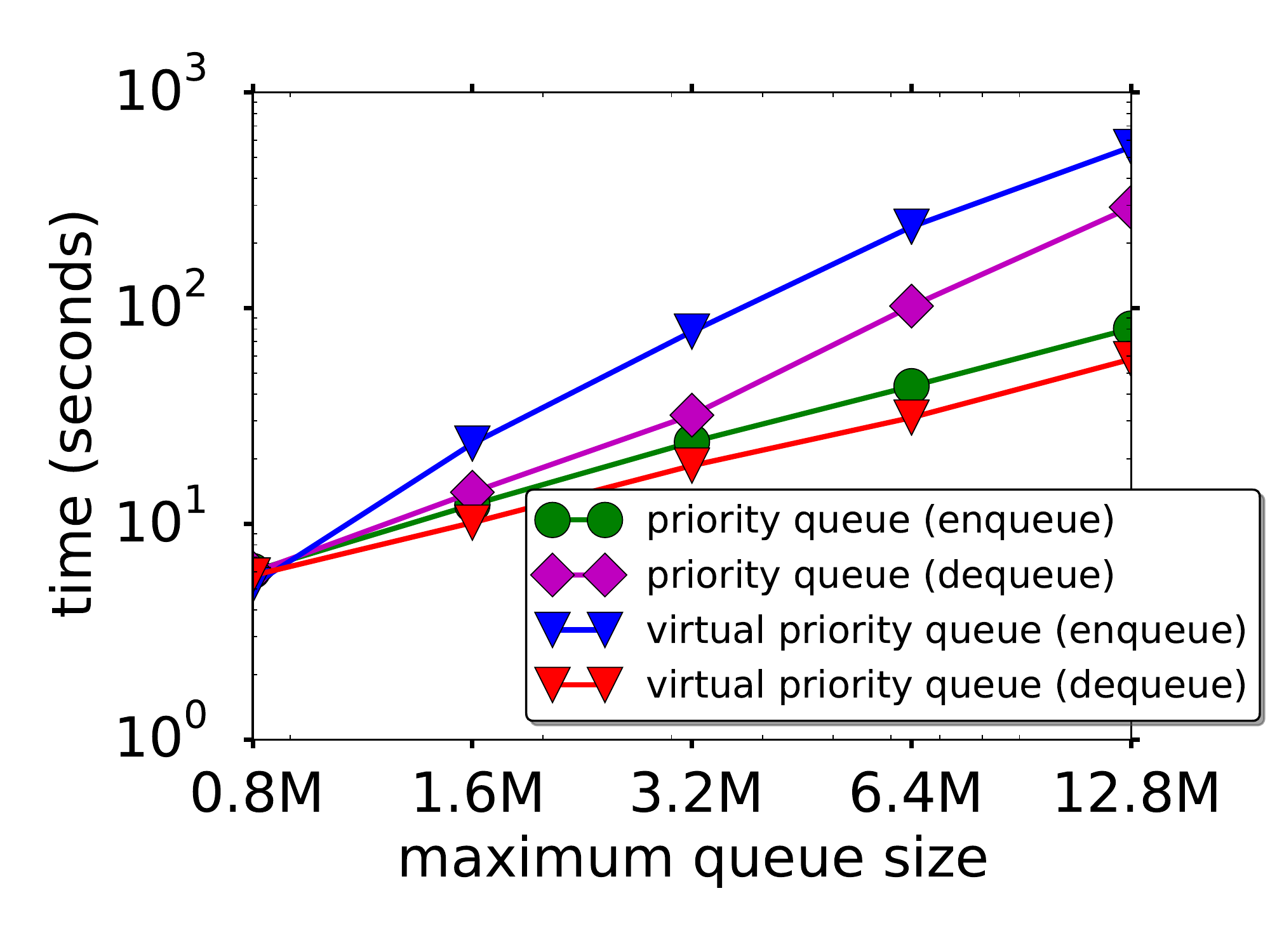}
    \caption{Virtual Priority Queue Result}\label{fig:queue_result}
\end{minipage}

\end{figure*}

\subsection{Effect of the Result Set Size ({\large  $\boldsymbol{k}$})}
\label{subsec:result_k}

The previous evaluations focused on top-$1$ computations.
As the result set size ($k$) increases, \oursyst tends to explore more subgraphs since it can start pruning out subgraphs only after its result set contains $k$ entries (i.e., as the result set needs to include more entries, \oursyst can prune out fewer subgraphs).
We measured the effect of $k$ on the number of candidate subgraphs examined by \oursyst as well as the overall time spent for subgraph discovery computations.
In this evaluation, we performed clique discovery on the \texttt{Email} dataset and pattern mining and subgraph isomorphism computations on the \texttt{CiteSeer} dataset. In Figure~\ref{fig:topkeffect}, as long as $k$ is smaller than $1000$, both the number of candidate subgraphs and running time vary insignificantly.
When $k$ is greater than $1000$, the number of candidate subgraphs and running time increase modestly with $k$.

\subsection{Performance of Virtual Priority Queue}
\label{subsec:result_queue}

We measured the performance of our virtual priority queue (Section~\ref{sec:architecture}) while configuring it to store subgraphs on an HGST Travelstar 2.5-Inch 1TB 5400RPM SATA 6Gbps 8MB Cache internal hard drive.
To carry out the measurement solely focusing on the enqueue and dequeue costs without being affected by individual subgraph computations, we first enqueued a certain number (0.8M, 1.6M, 3.2M, 6.4M, and 12.8M) of distinct 10-edge subgraphs (growing phase) and then dequeued all of them (shrinking phase).
In Figure~\ref{fig:queue_result}, curves labeled ``priority queue (enqueue)'' and ``priority queue (dequeue)'' show the duration of the growing and shrinking phases, respectively, when Java's  standard priority queue implementation (\texttt{PriorityQueue}) was used (up to 40GB of memory was required).
Curves labeled ``virtual priority queue (enqueue)'' and ``virtual priority queue (dequeue)'' show the time results when our virtual priority queue used at most 3GB of memory while storing on disk subgraphs that were not able to fit into the memory.

The above results show that, compared to \texttt{PriorityQueue} which manages all of the subgraphs in memory incurring much higher memory overheard, our virtual priority queue performs quite competitively (time costs are at most 1.8 times higher when both the growing and shrinking phases are considered, at most 7 times higher during the growing phase, and as low as 19\% of \texttt{PriorityQueue}'s dequeue cost during the shrinking phase).
The reason for our virtual priority queue's high enqueue and low dequeue costs is that it creates sorted runs during the growing phase (paying relatively high costs) and then significantly benefits from these sorted runs during the dequeue phase.


\section{Related Work}
\label{sec:related_work}
This section summarizes related work focusing on top-$k$ subgraph discovery and top-$k$ query processing.  
Other closely related work, particularly subgraph discovery computations as well as previous graph processing and subgraph discovery systems are extensively discussed in Sections~\ref{subsec:queries} and \ref{subsec:limitations}, respectively.

\noindent
{\bf Maximum Clique Discovery.}
Our maximum clique discovery implementation (Sections~\ref{subsec:non-aggregate} and \ref{subsec:api_noAgg}) is based on the CP algorithm~\cite{carraghan.orl1990.clique} which calculates, for each clique $s$, an upper bound on the size of the cliques that $s$ can expand into and then prunes out $s$ if its size upper bound is smaller than the size of largest clique(s) discovered.
Researchers have also developed algorithms that can more efficiently find maximum cliques by finding a tighter upper bound than CP~\cite{bron.commun73.clique,uno,eppstein.sea11.clique,cheng.sigkdd12.cliqueMPI}.
We leave the implementation of these algorithms for \oursyst as our future work.

\noindent
{\bf Top-$\boldsymbol{k}$ Pattern Mining.}
Given a single large graph, Grami~\cite{elseidy.pvldb14.grami} finds the patterns in the graph that are as frequent as the given threshold {\em without} ranking the patterns based on their frequency.
While our top-$k$ pattern mining implementation finds the $k$ most frequent patterns of a certain size (Sections~\ref{subsec:aggregate} and \ref{subsec:api_agg}), the closest work that we are aware of~\cite{zhu.pvldb11.topkfsm} finds the $k$ largest patterns that are as frequent as a given threshold.

\noindent
{\bf Top-$\boldsymbol{k}$ Subgraph Isomorphism.}
Our top-$k$ subgraph isomorphism implementation is motivated by Gupta et al.'s work~\cite{gupta.icde14.topksubgraph} which uses an index to quickly identify subgraphs whose expansions cannot produce any of the desired subgraphs (i.e., $k$ highest-scored subgraphs that match a query graph).
Zou et al. developed another solution that uses a different indexing approach~\cite{zou2007top}.
We intend to implement and evaluate this solution for \oursyst.

\noindent
{\bf Custom Top-$\boldsymbol{k}$ Subgraph Discovery.}
There are also subgraph discovery techniques that are designed to quickly obtain the $k$ subgraphs of highest preference according to a user-specified criterion~\cite{wu2013ontology,yang2016fast,hong2015subgraph,vasilyeva2014top,gupta.icde14.topksubgraph, macropol2010scalable, Bogdanov2013cliques}.
Our future work includes implementation of these techniques for \oursyst.

\noindent
{\bf Top-$\boldsymbol{k}$ Query Processing.}
In the context of database systems, various techniques for top-$k$ queries have been developed~\cite{ilyas.cs08.topK}.
These techniques commonly maintain (1) a top-$k$ result set, from which a threshold (the lowest-priority item in the set) is obtained, and (2) an upper bound on the priorities of unexamined data items.
If the threshold is higher than the upper bound (i.e., none of the unexamined data items can be included in the result set), the result set is returned as the final answer.
The top-$k$ query processing techniques for database systems cannot adequately support subgraph discovery computations since they are mainly designed for queries on relations rather than a large collection of subgraphs that need to be expanded according to their priorities.

\section{Conclusions}
\label{sec:conclusion}

We presented \oursyst, a new system for efficient top-$k$ subgraph discovery in large graphs.
\oursyst's API allows users to specify application-specific criteria for exploring and prioritizing subgraphs.
\oursyst's execution proceeds by creating unit subgraphs containing a vertex or an edge, and then efficiently discovering the result set by selectively expanding subgraphs according to their priority (prioritized expansion).
It also proactively discards subgraphs from which the desired subgraphs cannot be obtained (pruning).
For discovery computations with high space overhead, it provides efficient on-disk management of subgraphs. 

We evaluated \oursyst on real-world datasets of various sizes for three example computations: maximum clique discovery, subgraph isomorphism search, and pattern mining. 
\oursyst consistently outperformed the closest state-of-the-art alternative, achieving at least $2$ orders of magnitude improvement for clique discovery and $1$ order of magnitude improvement for subgraph isomorphism search and pattern mining, while utilizing $1/40$ of the computational resources compared to the closest alternative.

\bibliographystyle{abbrv}

\end{document}